\documentclass[10pt,a4paper]{article}
\usepackage[utf8]{inputenc}
\usepackage[english]{babel}
\usepackage{graphpap,amsmath,amsfonts,amssymb,epsfig,makeidx,graphicx,textcomp,cite,citeall} 
\usepackage[left=1.5cm, right=1.5cm, top=1.5cm, bottom=1.5cm]{geometry}

\begin{document}
	\begin{center}
		\Large\textbf{Thermodynamic behaviour of magnetocaloric quantities in spin-1/2
			Ising square trilayer}
	\end{center}
	
	\begin{center}
		{Soham Chandra}\\
		{\it Department of Physics, Presidency University}\\
		{\it 86/1 College Street, Calcutta-700073, India}\\
		{E-mail: soham.rs@presiuniv.ac.in}
	\end{center}
	\vspace{10pt}
	
	\begin{abstract}
		A spin-1/2, Ising trilayered ferrimagnetic system on square Bravais lattice is studied, employing Monte-Carlo simulation with the single spin-flip Metropolis algorithm. The bulk of such a system is formed by three layers, each of which is composed entirely either by A or B type of atoms, resulting in two distinct compositions: ABA and AAB and two different types of magnetic interactions: ferromagnetic between like atoms and antiferromagnetic between unlike atoms. For such systems, Inverse Absolute of Reduced Residual Magnetisation is the absolute value of the ratio of the extremum of the magnetisation in between compensation and critical points and the saturation magnetisation. Variation of relative interaction strengths in the Hamiltonian, for a range of values, leads to the shift of compensation point and critical point and changes in the magnitude of Inverse absolute of Reduced Residual magnetisation. Probable mathematical forms of dependences of the Inverse absolute of Redced Residual magnetisation and temperature interval between the compensation and critical points on controlling parameters were proposed in the absence of applied magnetic field and have obtained phase diagrams for both types of configurations from these relations. This alternative description of the simulated systems may help technologists design magnetocaloric materials according to desired characteristics.
	\end{abstract}
\begin{figure}[!htb]
	\begin{center}
		\begin{tabular}{c}
			(a)
			\resizebox{8.5cm}{!}{\includegraphics[angle=0]{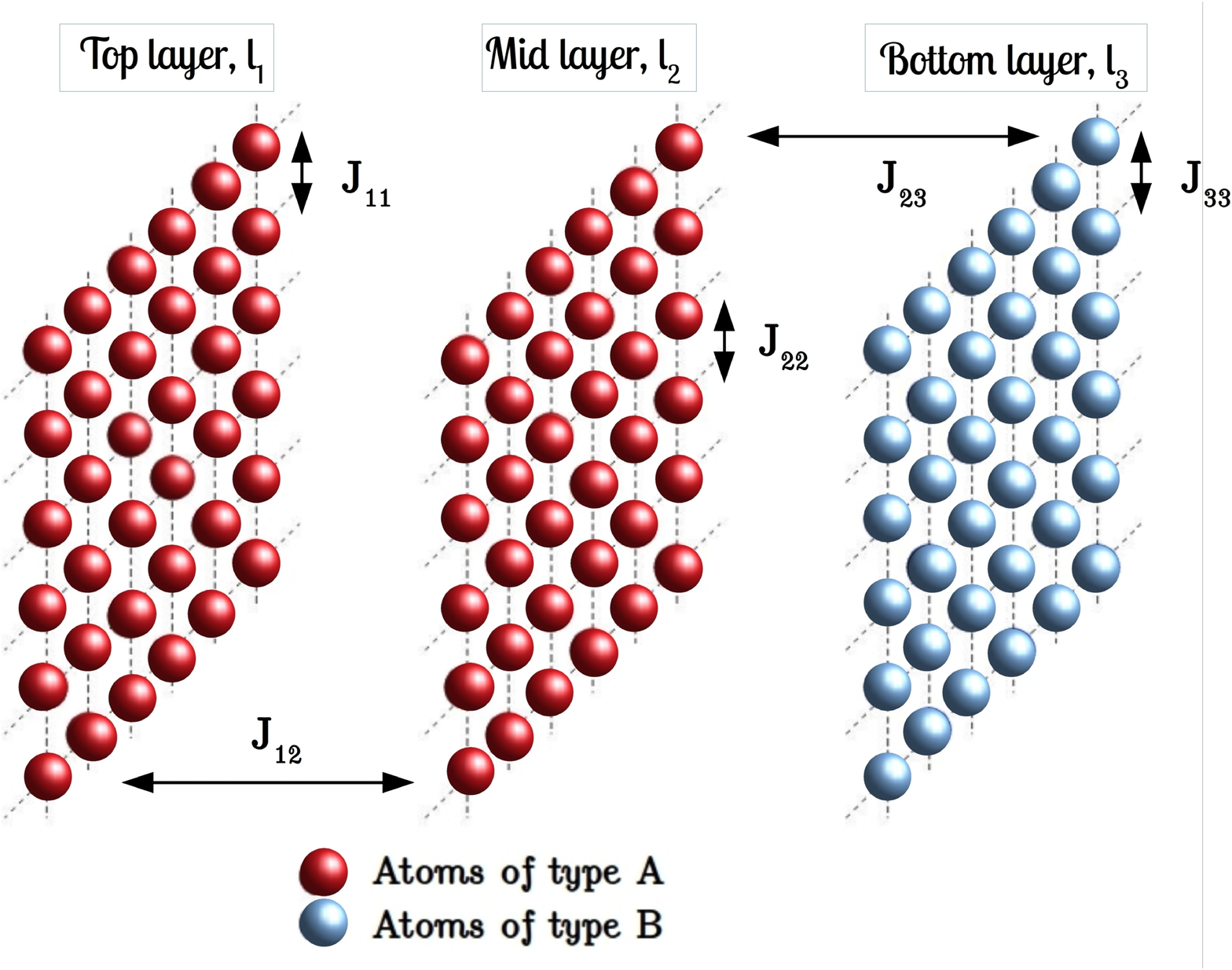}}
			(b)
			\resizebox{8.5cm}{!}{\includegraphics[angle=0]{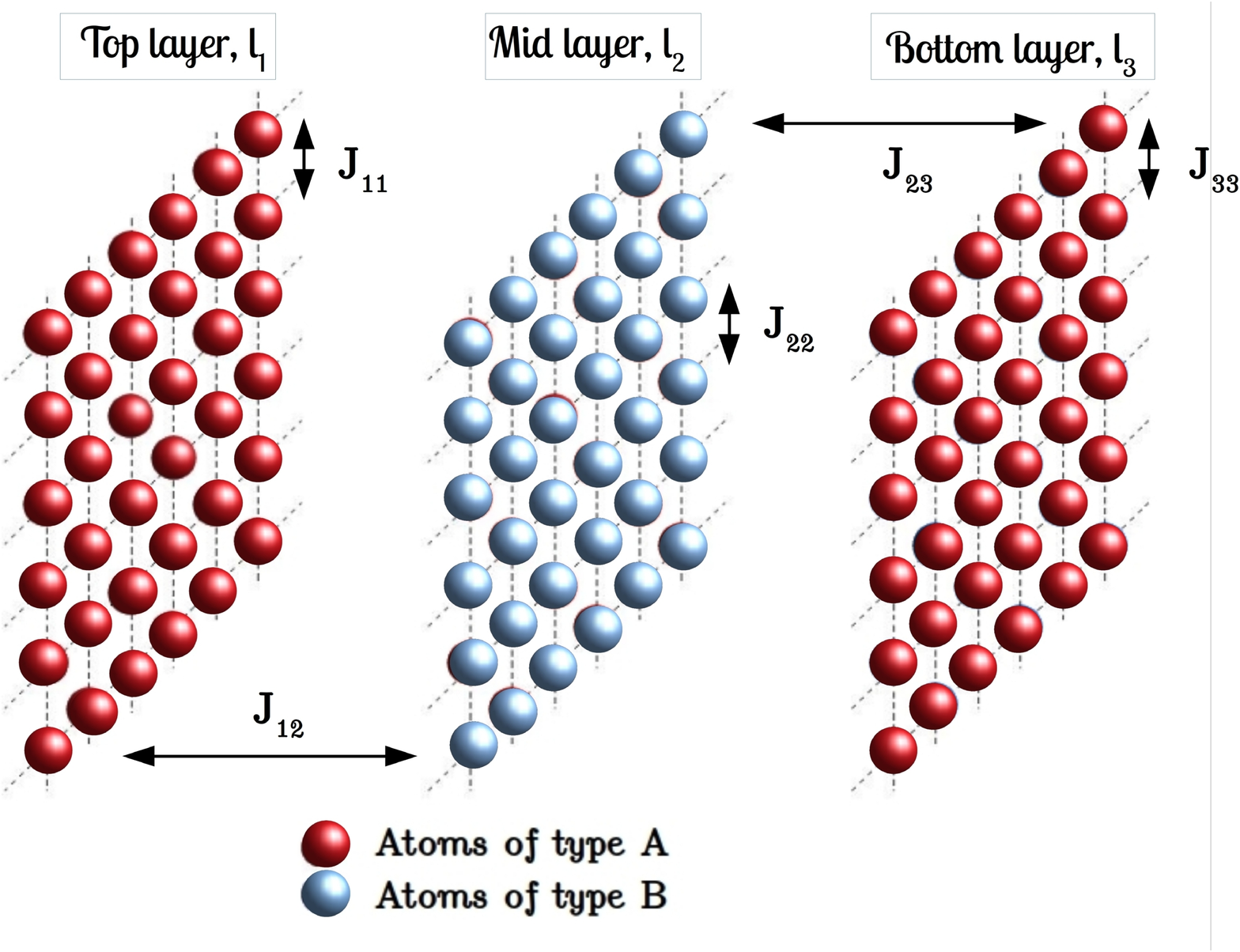}}
		\end{tabular}
		\caption{Two distinct trilayer stackings: (a) AAB stacking (with $J_{11}=J_{22}=J_{12}=J_{AA}$, $J_{33}=J_{BB}$ and $J_{23}=J_{AB}$)(b) ABA stacking (with $J_{11}=J_{33}=J_{AA}$, $J_{22}=J_{BB}$ and $J_{12}=J_{23}=J_{AB}$).}
		\label{fig_lattice_structure}
	\end{center}
\end{figure}

\vspace{5pt}
\newpage
\begin{center} {\Large \textbf {I. Introduction}}\end{center}
Ferrimagnetism was discovered in 1948 \cite{Cullity}. In the past decades, studies on ferrimagnets have revealed unique properties, showing their potential in technological applications like magneto-optical devices \cite{Connell, Ostorero}, Giant magnetoresistance (GMR) \cite{Camley} and, Magnetocaloric effect (MCE) \cite{Phan, Ma} etc. A ferrimagnet is modelled as a combination of two or more magnetic substructures, like, subsets of atoms, sublattices, or layers. Layered ferrimagnetic materials, owing to their enhanced surface-to-volume ratio, present characteristics quite different from the bulk as the surface atoms may provide significant contributions to the physical properties of the system, as a whole. With the advent of different thin film growth techniques, like, molecular-beam epitaxy (MBE) \cite{Herman}, atomic layer deposition (ALD) \cite{George}, metalorganic chemical vapor deposition (MOCVD) \cite{Stringfellow}, and pulsed laser deposition (PLD) \cite{Singh1}, experimental growths of bilayered \cite{Stier}, trilayered \cite{Smits}, and multilayered \cite{Chern,Sankowski,Maitra} systems are realized with desired characteristics. It is one of the reasons behind the
increased interest in the theoretical and experimental studies of magnetic properties of layered ferrimagnets.\\
\indent For a layered ferrimagnet, each of the layers may have different thermal dependencies for magnetization. Such non-identical behaviours, when combined, may show some interesting phenomenon such as \textit{compensation}, i.e., there exists temperature below the critical point, called \textbf{compensation point}, for which total magnetization of the bulk vanishes while individual layers have magnetic order \cite{Cullity}. Compensation is not related to criticality of the system and physical properties like the magnetic coercivity exhibit singular behaviour at the compensation point \cite{Connell, Ostorero}. Compensation point of some ferrimagnetic materials at room temperature and strong temperature dependence of coercive field around compensation point makes these materials particularly useful for thermomagnetic recording devices \cite{Connell}. In 2D Ising planar ferrimagnets, compensation occurs in mixed-spin cases with combinations of different spins \cite{Godoy1, Nakamura, Godoy2}. For layered single-spin systems having even number of layers, the necessity of site dilution and strict restrictions on the conditions of controlling parameters for the occurrence of compensation effect has been verified through pair approximation (PA) calculations \cite{Balcerzak2, Szalowski, Szalowski2} for the Ising-Heisenberg systems and by MC simulations for the Ising systems \cite{Diaz1, Diaz2}. But for an odd number of layers, neither cite dilution \cite{Santos} nor mixed-spin cases \cite{Diaz3, Diaz4} is necessary for observing compensation. That is why the area of interest in this article is the compensation point, $T_{comp}$, of the trilayered spin-1/2 Ising ferrimagnet. For being a single-spin model, such systems are, computationally and growth-wise, less expensive than other ferrimagnetic models and belong to the simplest of spin-systems for compensation. In literature, using Ising mechanics, layered systems have been studied by equilibrium Monte Carlo (MC) simulations in \cite{Laosiritaworn, Albano}, by mean-field approximation (MFA) in \cite{Lubensky}, by effective-field approximation (EFA) in \cite{Kaneyoshi1, Kaneyoshi2, Kaneyoshi3, Kaneyoshi4}, by series-expansion method in \cite{Oitmaa}, by renormalization-group (RG) method in \cite{Ohno}, by spin-fluctuation theory in \cite{Benneman}, by exact recursion equation on the Bethe lattice in \cite{Albayrak} and by cluster variation method in pair approximation in \cite{Balcerzak1}.\\
\indent \textit{Magnetocaloric effect} (MCE) is defined by heating or cooling of a magnetic material, with the variation of the applied magnetic field. Thus it is a suitable candidate for a new type of refrigerant materials for construction of energy-efficient devices and that is why its potential is investigated in \cite{Spichkin, Gschneidner}. After the discovery of MCE in iron by Warburg in 1881 \cite{Warburg}, Debye in 1926 \cite{Debye} and Giauque in 1927 \cite{Giauque} provided theoretical explanations. Magnetic entropy change, $\Delta S$ and/or adiabatic temperature change, $\Delta T$ characterise MCE and the Maxwell's relation $\left(\dfrac{\partial S}{\partial H}\right)_{T}=\left(\dfrac{\partial M}{\partial T}\right)_{H}$ connects them, with $H$ and $T$ are the applied magnetic field and the temperature of the system, respectively. This relation shows, for an abrupt change in magnetisation around the compensation point, for most of the ferrimagnets with compensation, large MCE may be expected. For the first-order phase transitions, a large change in entropy due to the sharp change in magnetization usually is observed in the neighbourhood of transition \cite{Pecharsky, Tegus}. But such materials suffer from hysteretic behaviour and a narrow range of working temperature, as a magnetic refrigerant \cite{Provenzano}. But, second-order transition materials, because of the absence of magnetic and thermal hysteresis and wide interval of transition temperatures, are extensively studied \cite{Xie}. At present most of them are operated near their transition temperatures for magnetic refrigeration. In layered ferrimagnets, especially with an odd number of layers, across compensation point, a sharp change in magnetization is observed. Thus such materials have an advantage over conventional second-order magnetic transition materials, in having a lower temperature than the critical point with a high value of $\Delta M$. But such ferrimagnetic materials, yet have not been widely studied as magnetocaloric materials.\\
\indent Because of a limited number of exact methods available, approximate and numerical studies have their own importance and a brief referral to some such recent studies on compensation, in the literature, follows. It is reported, by MFA and EFA in \cite{Diaz3} and by MC simulations with Single cluster Wolff Algorithm in \cite{Diaz4} that under a threshold of different types of interaction strengths between lattice sites, unlike temperature dependencies of sublattice magnetisations cause the compensation point to appear in an Ising trilayer on a square lattice. A quasi three-dimensional, spin-1/2, Ising trilayer stacking on a square lattice with quenched non-magnetic impurity is studied by MC simulation in \cite{Sajid}. There it was observed, the critical and compensation points drift towards lower temperatures with increasing concentration of non-magnetic impurities. It was suggested in \cite{Soham} that Inverse Absolute Reduced Residual Magnetisation (IARRM), another magnetocaloric quantity possibly varies systematically. The dilution effects on compensation temperature were studied in \cite{Fadil1}, in a nanotrilayer graphene structure through MC simulation and observed that with the reduction in dilution probability, the compensation temperature, \textit{increases} in an ABA system and \textit{decreases} in BAB systems. In \cite{Fadil2}, by the MC approach, the appearance of two compensation temperatures is reported in a mixed spin $(7/2,1)$ antiferromagnetic ovalene nanostructures. By MC simulation \cite{Fadil3}, in Blume Capel model of a bilayer graphene structure with Ruderman-Kittel-Kasuya-Yosida (RKKY) interactions, the transition temperature was observed to increase with a decrease in the number of non-magnetic layers. Dilution effects on compensation temperature in borophene core-shell structure are studied in \cite{Fadil4} by MC simulations and analysed the existence of the compensation and transition temperatures for unique physical parameters and hysteresis cycles. The trilayered Blume-Capel $(S=1)$ magnet is studied in \cite{Muktish} by MC simulation scheme and the equilibrium behaviours of critical and compensation temperatures and reported the dependence of these two temperatures on the anisotropy and obtained comprehensive phase diagrams in Hamiltonian parameter space.\\
\indent Inverse Absolute of Reduced Residual Magnetisation (IARRM, for brevity), defined by the absolute value of the ratio of the extremum value of the magnetisation in between compensation and critical points and the saturation magnetisation, is a novel magnetocaloric quantity of interest. In this study, variation of IARRM against controlling parameters is observed. The objective is to find possible mathematical forms of dependences of two unconventional quantities, one being the IARRM, on the controlling parameters namely relative interaction strengths which might be helpful for experimentalists in designing materials for their desired magnetocaloric characteristics.\\
\indent The formation of this article is as follows. Details of the microscopic structure and the Ising Hamiltonian for the trilayer system are in Section II. In Section III, the MC simulation scheme and the calculated quantities are discussed. Numerical results are presented in Section IV. Next, in Section V, there are concluding remarks and summary of the article is in Section VI.
\vspace{20pt}
\begin{center} {\Large \textbf {II. Model}}\end{center}
\indent In this study, an Ising superlattice containing three magnetic layers on square lattice with the following details (as in \cite{Diaz3,Diaz4}) is considered:
\begin{itemize}
	\item[(a)] Each layer is exhaustively composed by either of the two possible types of atoms, A or B with the coordination number being 6 or 5 depending on that site being in the mid-layer or in the surface layers respectively.  
	\item[(b)] Three different interactions of two types, between atoms, exist:\\
	A-A $\to$ Ferromagnetic\\
	B-B $\to$ Ferromagnetic\\
	A-B $\to$ Anti-ferromagnetic
	\item[(c)] Two distinct stackings are possible: (i) AAB [Fig.-\ref{fig_lattice_structure}a] and (ii) ABA [Fig.-\ref{fig_lattice_structure}b].	 
\end{itemize}

\begin{figure}[!htb]
	\begin{center}
		\begin{tabular}{c}
			(a)
			\resizebox{8.0cm}{!}{\includegraphics[angle=0]{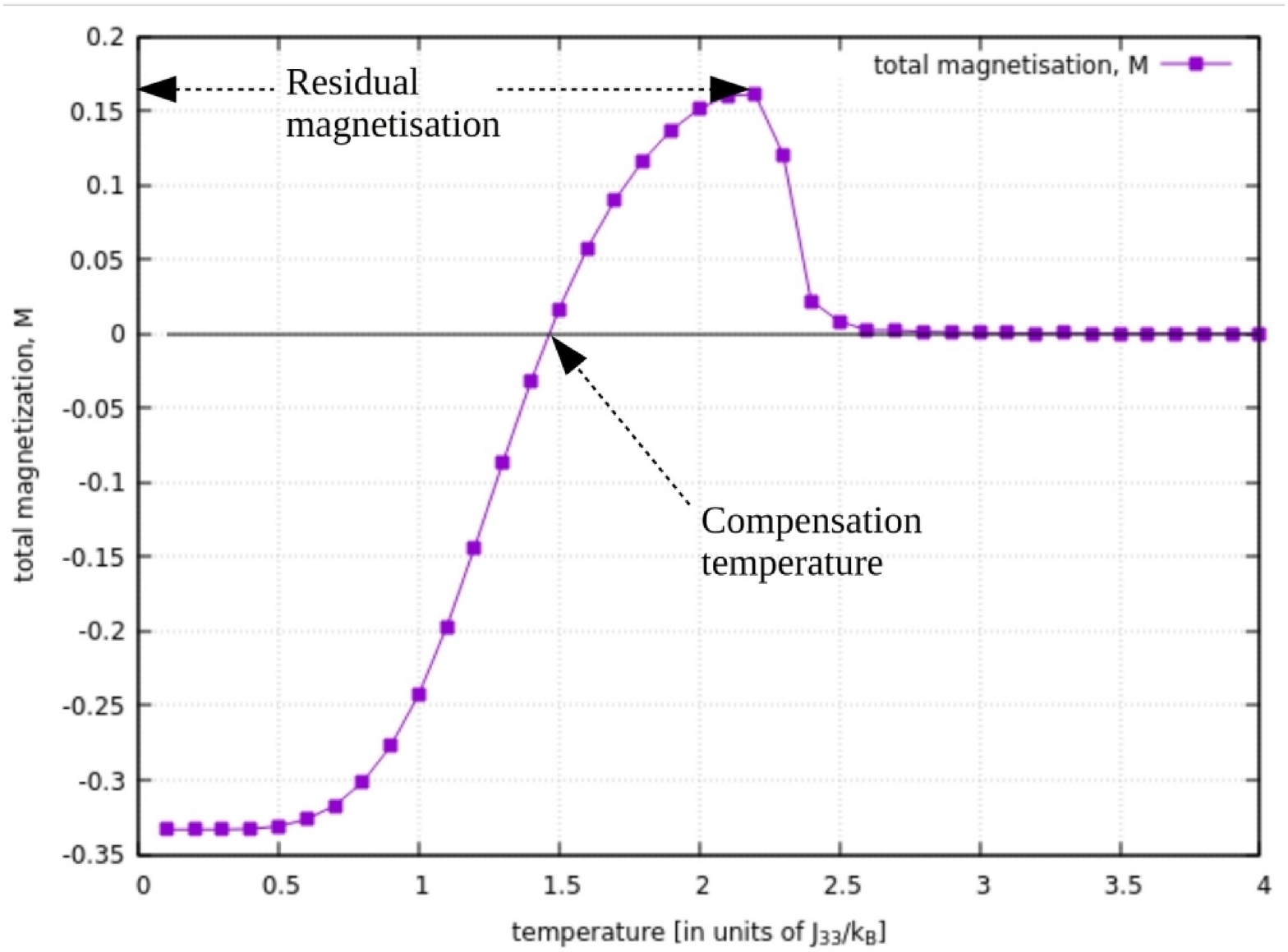}}
			(b)
			\resizebox{8.0cm}{!}{\includegraphics[angle=0]{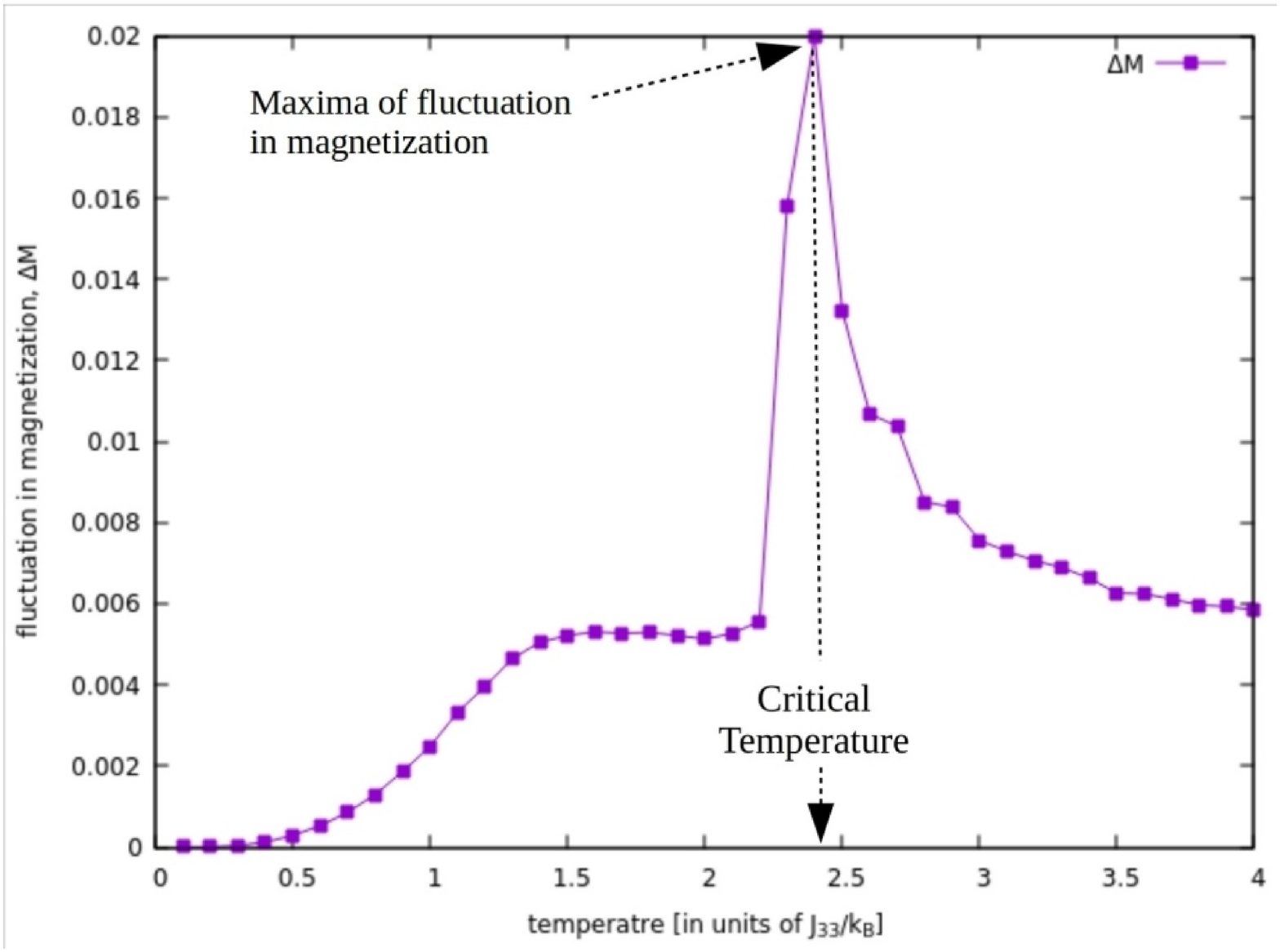}}
		\end{tabular}
		\caption{The plots of AAB cofiguration of (a) total magnetisation vs. temperature and (b) fluctuation in magnetization vs. temperature [for $J_{AA}/J_{BB}=0.3$ and $J_{AB}/J_{BB}=-0.4$] show how observed quantities are extracted}		
		\label{fig_aab_mag_fluc}
	\end{center}
\end{figure}

The equilibrium spin configuration on any layer is determined by minimising the free energy of the whole system. As the spins interact Ising-like, in-plane as well as out-of-plane, the Hamiltonian for the trilayer system is:
\begin{equation}
\label{eq_Hamiltonian}
H=-J_{11}\sum_{<t,{t}^{\prime }>}S_{t}S_{{t}^{\prime }}-J_{22}\sum_{<m,{m}^{\prime }>}S_{m}S_{{m}^{\prime }}-J_{33}\sum_{<b,{b}^{\prime }>}S_{b}S_{{b}^{\prime }}-J_{12}\sum_{<t,m>}S_{t}S_{m}-J_{23}\sum_{<m,b>}S_{m}S_{b}
\end{equation}
where $\langle t,{t}^{\prime }\rangle$, $\langle m,{m}^{\prime }\rangle$, $\langle b,{b}^{\prime }\rangle$ are, respectively, summations over all nearest-neighbor pairs in the top, mid and bottom layers and $\langle t,m\rangle$, $\langle m,b\rangle$ are, respectively, summations over pairs of nearest-neighbor sites in adjacent layers, top \& middle and middle \& bottom layers. In Equation [\ref{eq_Hamiltonian}], the first, second and third terms respectively are for the intra-planar contributions from the top-layer, mid-layer and bottom-layer. The fourth and the fifth terms arise out of the nearest neighbour inter-planar interactions, between top and mid layers and mid and bottom layers.\\
\indent For the AAB type trilayer system, the nature of the coupling strengths in Equation [\ref{eq_Hamiltonian}] are: $J_{11}>0$ , $J_{22}>0$, $J_{33}>0$ and $J_{12}>0$, $J_{23}<0$ and it is evident, $J_{11}=J_{22}=J_{12}=J_{AA}$, $J_{33}=J_{BB}$ and $J_{23}=J_{AB}$. As there is no spacer within the layers, the interatomic A-A distance remains equal to their equilibrium separation, in-plane and out-of-plane. So for the AAB configuration, $J_{12}=J_{AA}$.\\
\indent But when switched to ABA type system, we have in Equation [\ref{eq_Hamiltonian}]: $J_{11}>0$ , $J_{22}>0$, $J_{33}>0$ and $J_{12}<0$, $J_{23}<0$ and we call $J_{11}=J_{33}=J_{AA}$, $J_{22}=J_{BB}$ and $J_{12}=J_{23}=J_{AB}$. Periodic boundary conditions in-plane and open boundary conditions along the vertical are considered so that there is no out-of-plane interaction term between the top and bottom layer in the Hamiltonian.\\
\vspace{20pt}
\begin{center} {\Large \textbf {III. Simulation Scheme}}\end{center}
\indent The model described above was simulated using the Monte Carlo simulations with Metropolis single spin-flip algorithm \cite{Landau, Binder} with each plane having $L^{2}$ sites with a system size having $L=100$. For $L\geqslant60$ compensation point tends to a stable value \cite{Diaz4}, thus the lattice size in our study is quite standard. The initial high-temperature paramagnetic phase of spin configurations had randomly selected 50\% of the total number of spins in upward projection with $S_{i}=+1$ and the rest in downward projection with $S_{i}=-1$ (Using $1$ instead of $1/2$ rescales the coupling constants only).At a fixed temperature $T$, Metropolis rate \cite{Metropolis, Newman} governed the spin flipping from $S_{i}$ to $-S_{i}$:
\begin{equation}
\label{eq_metropolis}
P(S_{i} \to -S_{i}) = \text{min} \{1, \exp (-\Delta E/k_{B}T)\}
\end{equation}
where $\Delta E$ is the change in internal energy due to the $i$-th spin projection changing from $S_{i}$ to $-S_{i}$. Boltzmann constant, $k_{B}$, is set to $1$. Similar $3L^{2}$ random updates of spins constitute one Monte Carlo sweep (MCS) of the entire system and define one unit of time in our study. At each temperature step, starting from the configuration of the previous temperature, the system was equilibrated till $5\times10^{4}$ MCS (that is equivalent to the allowance of a long enough \textit{time}, for equilibration) and thermal averages were calculated from further $5\times10^{4}$ MCS, i.e. the total MCS, $N=10^{5}$.\\
\indent The systems were observed for ten equidistant values of $J_{AA}/J_{BB}$, starting from $0.1$ to $1.0$ and for each fixed value of $J_{AA}/J_{BB}$, $J_{AB}/J_{BB}$ was varied from $-0.1$ to $-1.0$ with an interval of $-0.1$. For each of the combinations of $J_{AA}/J_{BB}$ and $J_{AB}/J_{BB}$, the time (or, ensemble) averages of the following quantities at each of the temperature points, were calculated in the following manner:\\
\textbf{(1) Sublattice magnetisations} for top, mid and bottom layers calculated, identically, at say, $i$-th MCS after equilibration, denoted by $M_{qi}$, by:
\begin{equation}
M_{qi}=\frac{1}{L^{2}}\sum_{x,y=1}^{L} \left( S_{qi}\right)_{xy}
\end{equation}
and the sum extends over all sites in each of the planes as x and y denote the co-ordinates of a spin on a plane and runs from $1$ to $L$ (which is $100$, in our study). Then we get the time (or, ensemble) average, from the last $N/2$ MCS, as follows:
\begin{equation}
\langle M_{q}\rangle=\dfrac{2}{N}\sum_{i=\frac{N}{2}+1}^{N}M_{qi}
\end{equation} 
where $q$ is to be replaced by $t,m\text{ or }b$ for top, mid and bottom layers and  $\langle\cdots\rangle$ denotes a time average (equivalently ensemble average) after attaining equilibrium.\\
\textbf{(2)} Time average value of \textbf{Average magnetisation of the trilayer} by $\langle M\rangle=\dfrac{1}{3}\left(\langle M_{t}\rangle+\langle M_{m}\rangle+\langle M_{b}\rangle\right)$\\
\textbf{(3)} After attaining equilibrium, \textbf{fluctuation in magnetisation, $\Delta M$} was calculated from the final $N/2$ MCS as follows:
\begin{equation}
{\Delta M}=\sqrt{\dfrac{2}{N-2} \sum_{i=\frac{N}{2}+1}^{N} (M_{i}-\overline{M})^{2} }
\end{equation}
where $M_{i}$ is the value of magnetisation of the whole system, calculated after the completion of $i$-th MCS and $\overline{M}$ is the average value of total magnetisation calculated over the $N/2$ MCS after equilibration.\\
\indent All the obtained values were stored and then sample averages were made (over 10 statistically independent microscopic samples with same macroscopic initial conditions) to report the values of these quantities for further investigation.
\vspace{20pt}
\begin{center} {\Large \textbf {IV. Results and Discussion}}\end{center}
From the plot of Fig.-\ref{fig_aab_mag_fluc}a, the presence of compensation was noted first, and then the coordinates of immediate neighbouring points on both sides of zero line were used to find out the value of compensation point by the method of linear interpolation. Then, the value of \textit{Inverse Absolute of Reduced Residual Magnetisation} (IARRM, $\mu$ for brevity) were calculated, which is a dimensionless quantity, by calculating the absolute value of the ratio between intermediate maximum/minimum value of magnetization between critical and compensation temperature and the value of magnetization at the lowest simulational temperature ($\approx$ saturation magnetization).\\
From the plot of Figure \ref{fig_aab_mag_fluc}(b), the value of temperature where the fluctuation in magnetization reaches its maximum value was noted, and that temperature was quoted as the critical temperature. Subsequently, the difference between the critical and compensation temperatures ($\Delta T$ for brevity) were obtained. After that, the data obtained so far were fitted to get the possible functional forms for these two quantities. The best values of underlying variables were worked with to figure out qualitative mathematical dependences.

\begin{figure}[!htb]
	\begin{center}
		\begin{tabular}{c}
			(a)
			\resizebox{5.0cm}{!}{\includegraphics[angle=0]{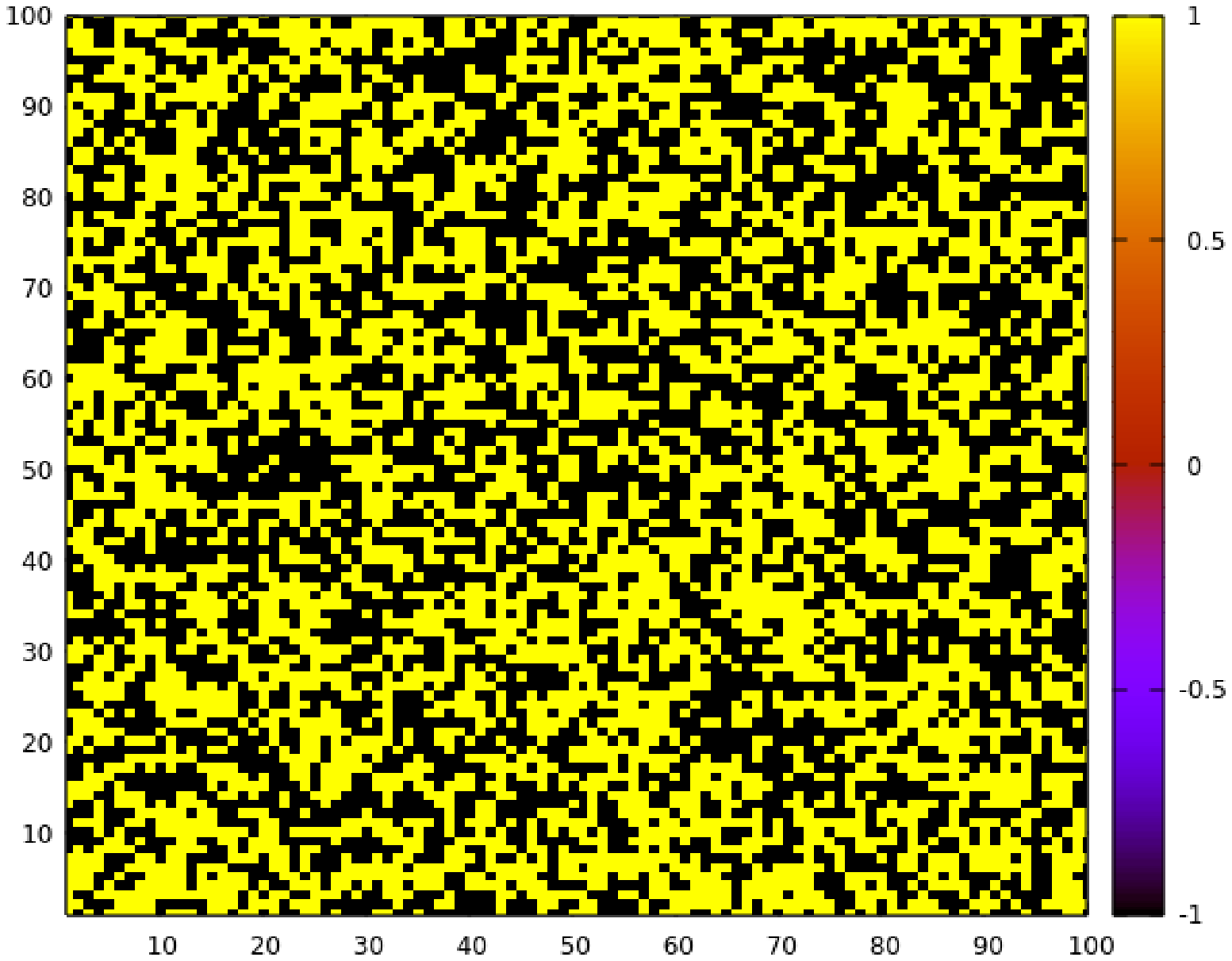}}
			(b)
			\resizebox{5.0cm}{!}{\includegraphics[angle=0]{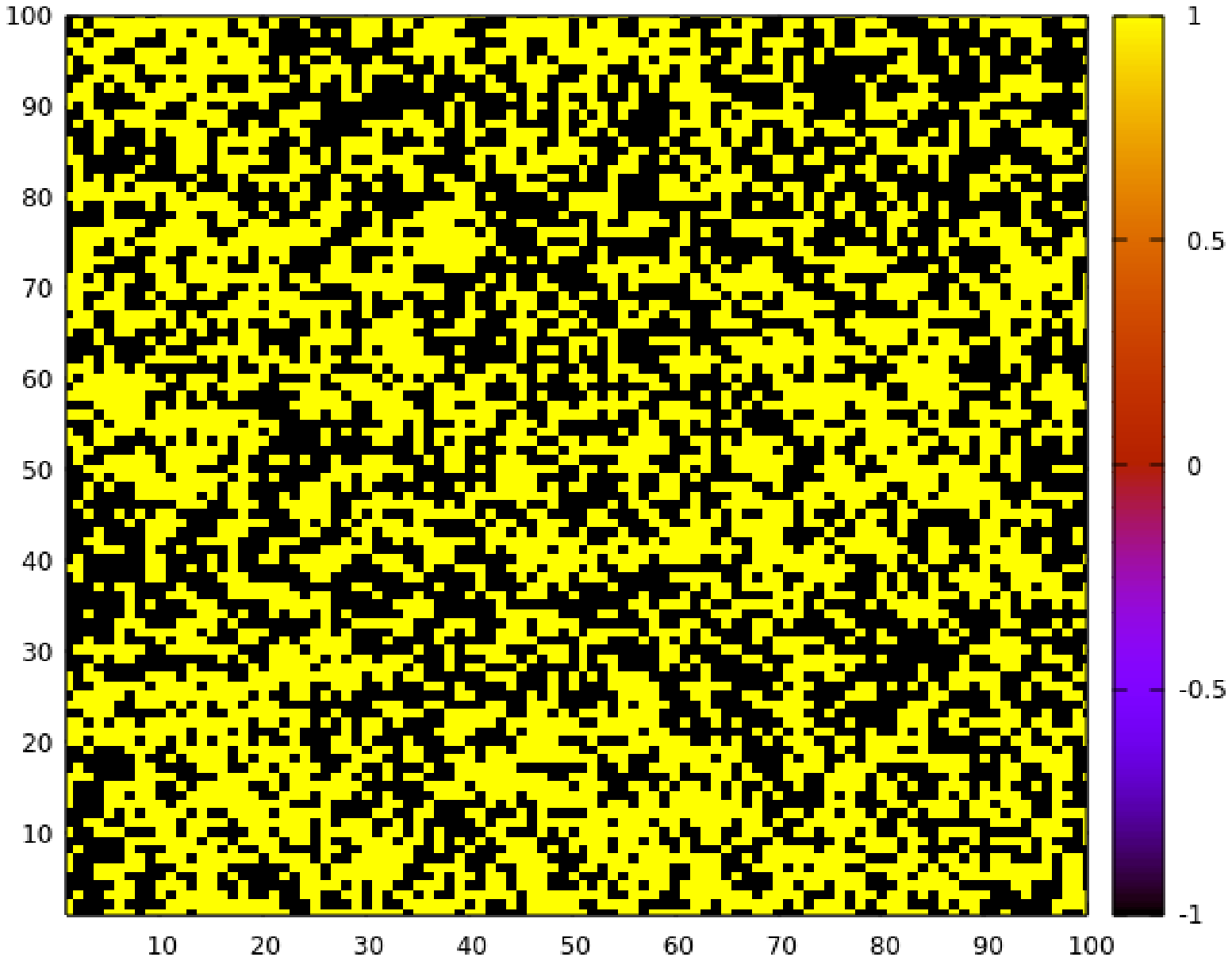}}
			(c)
			\resizebox{5.0cm}{!}{\includegraphics[angle=0]{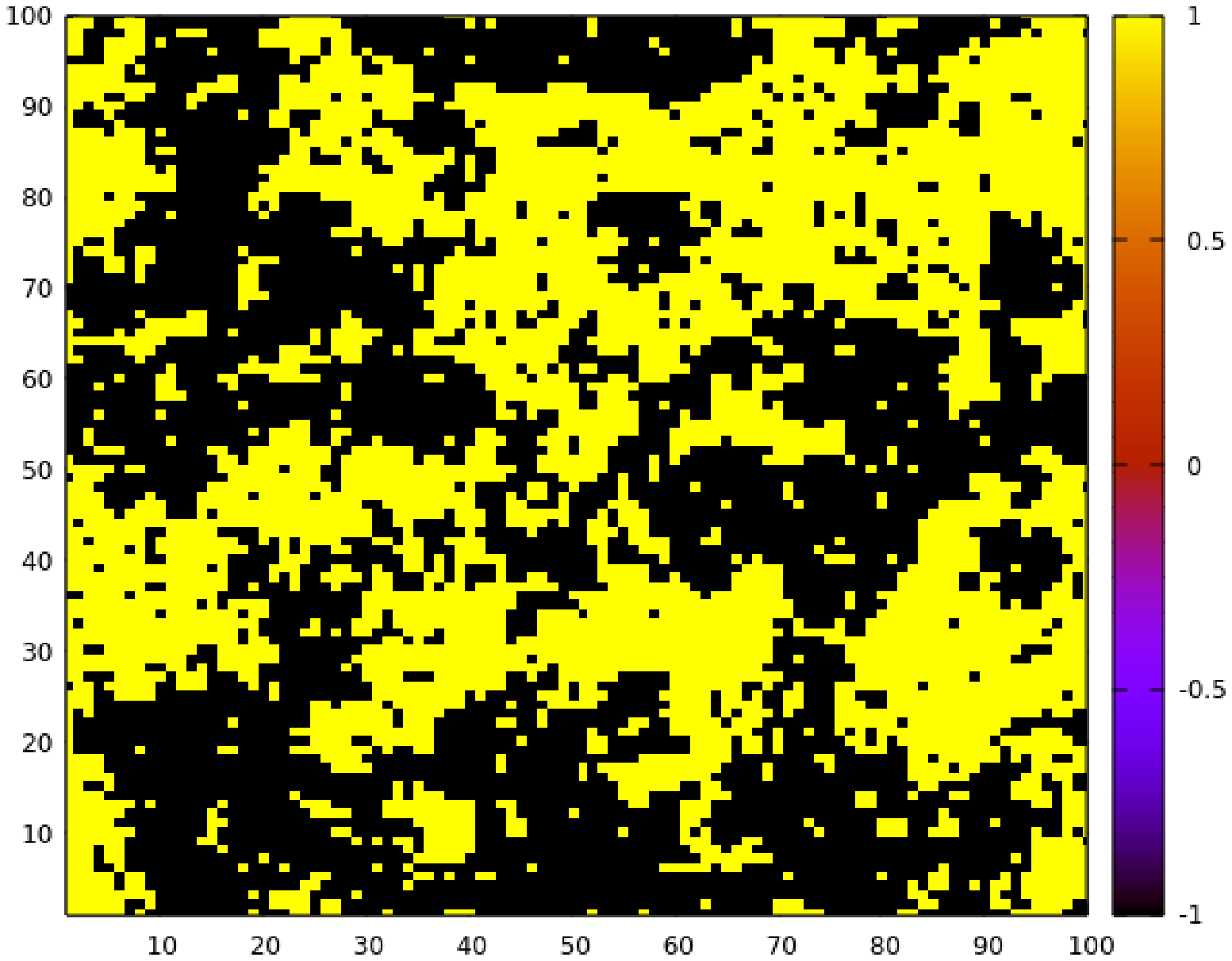}}
		\end{tabular}
		\caption{Morphology of (a) Top layer, (b) Mid layer and (c) Bottom layer for the AAB stacking ($J_{AA}/J_{BB}=0.3$ and $J_{AB}/J_{BB}=-0.4$) at $T_{crit}=2.4$ with $M_{t}=-0.01$, $M_{m}=-0.05$, $M_{b}=0.17$.}		
		\label{fig_aab_morpho1}
	\end{center}
\end{figure}

\begin{figure}[!htb]
	\begin{center}
		\begin{tabular}{c}
			(a)
			\resizebox{5.0cm}{!}{\includegraphics[angle=0]{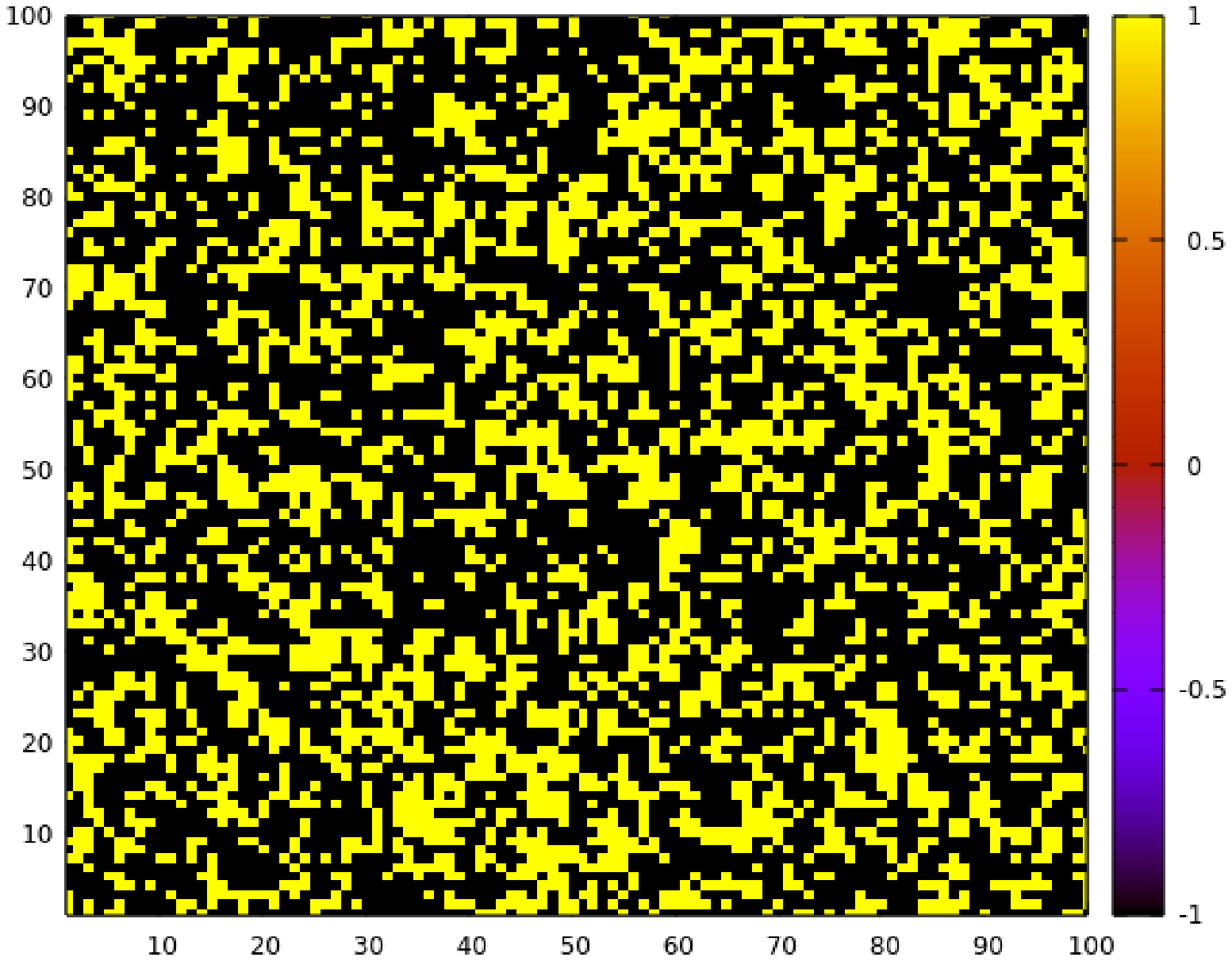}}
			(b)
			\resizebox{5.0cm}{!}{\includegraphics[angle=0]{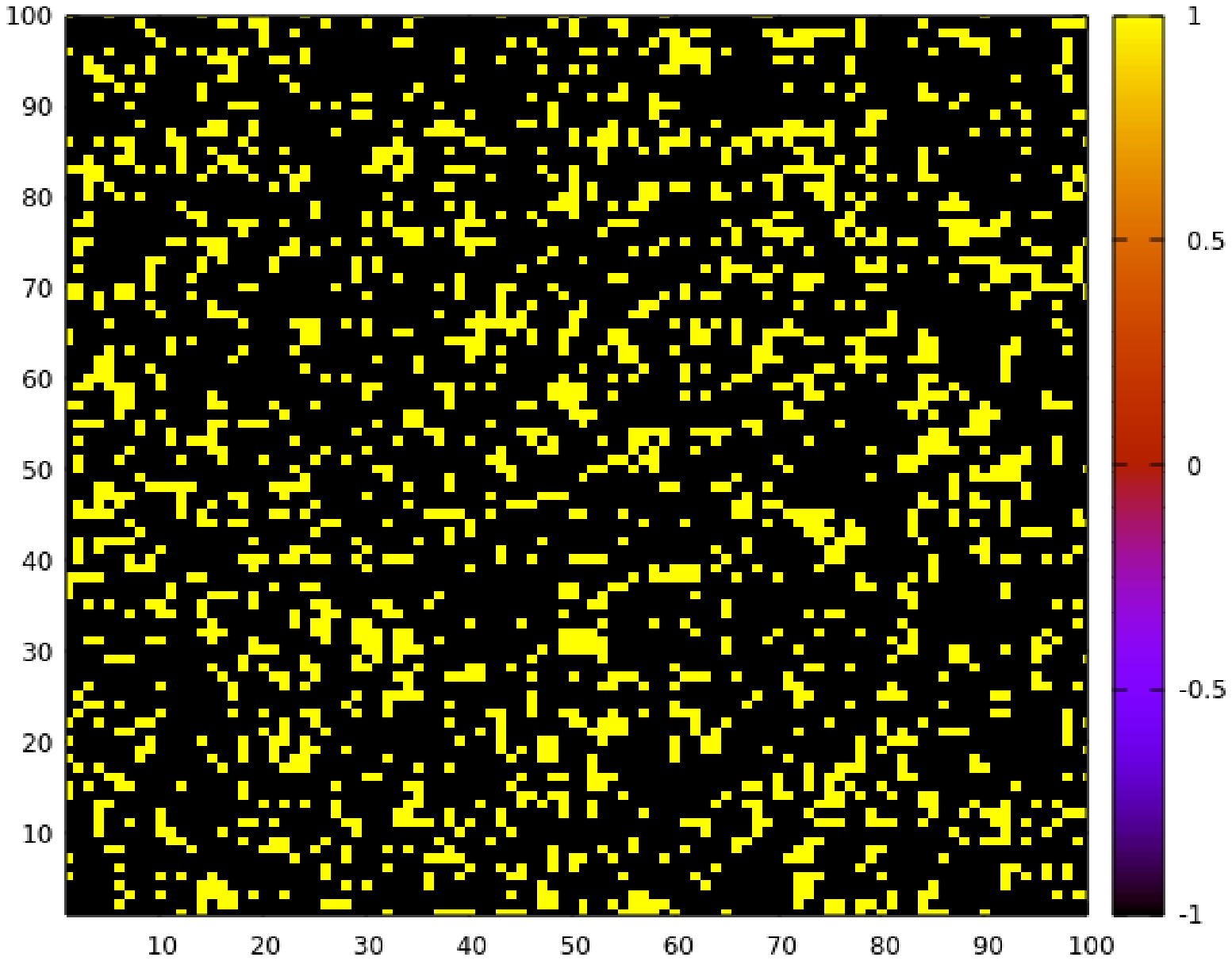}}
			(c)
			\resizebox{5.0cm}{!}{\includegraphics[angle=0]{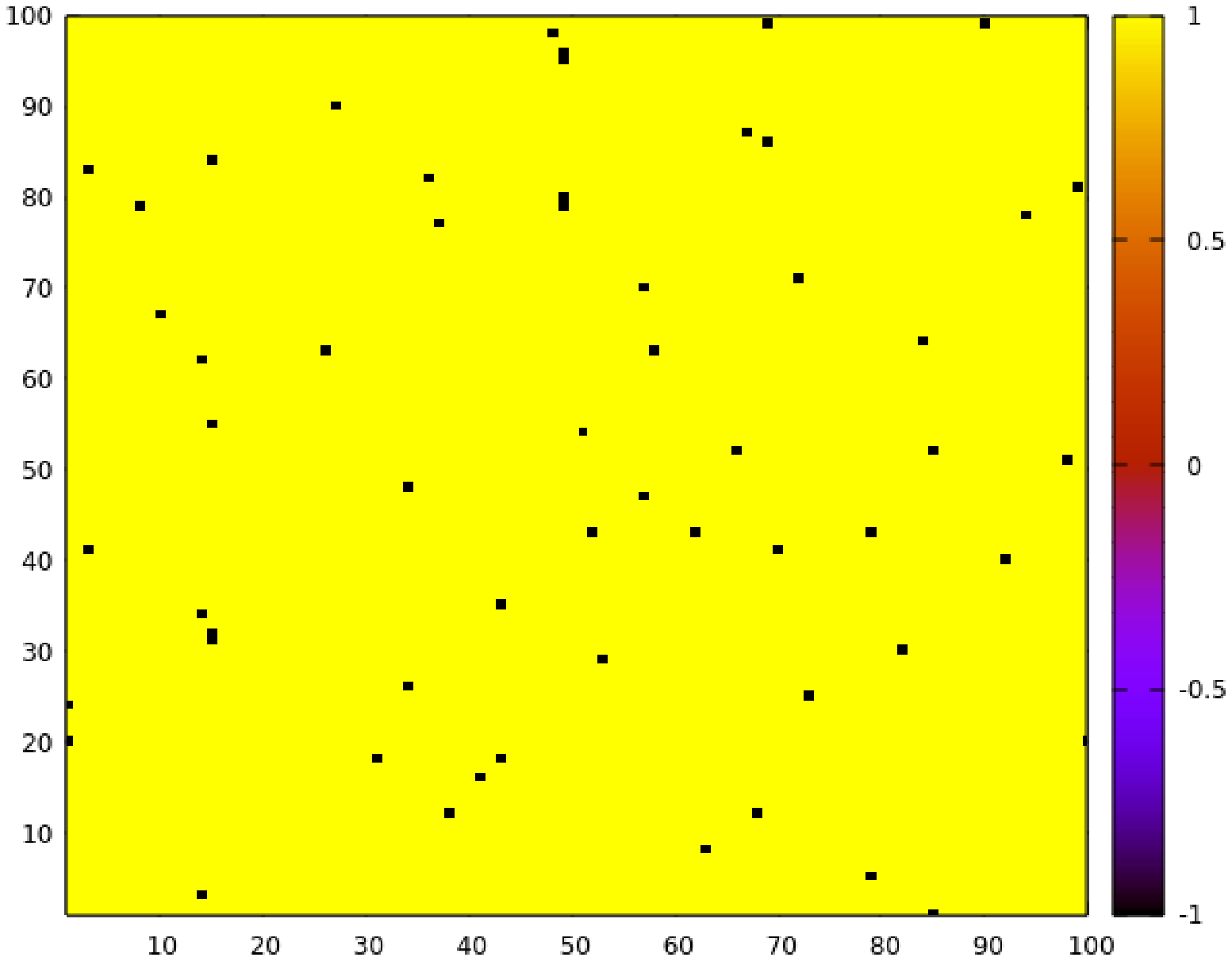}}
		\end{tabular}
		\caption{Morphology of (a) Top layer, (b) Mid layer and (c) Bottom layer for the AAB stacking ($J_{AA}/J_{BB}=0.3$ and $J_{AB}/J_{BB}=-0.4$) at $T=1.5$ (immediate higher neighbour than $T_{comp}$) with $M_{t}=-0.32$, $M_{m}=-0.62$, $M_{b}=0.99$.}		
		\label{fig_aab_morpho2}
	\end{center}
\end{figure}

\begin{figure}[!htb]
		\begin{center}
		\begin{tabular}{c}
			(a)
			\resizebox{5.0cm}{!}{\includegraphics[angle=0]{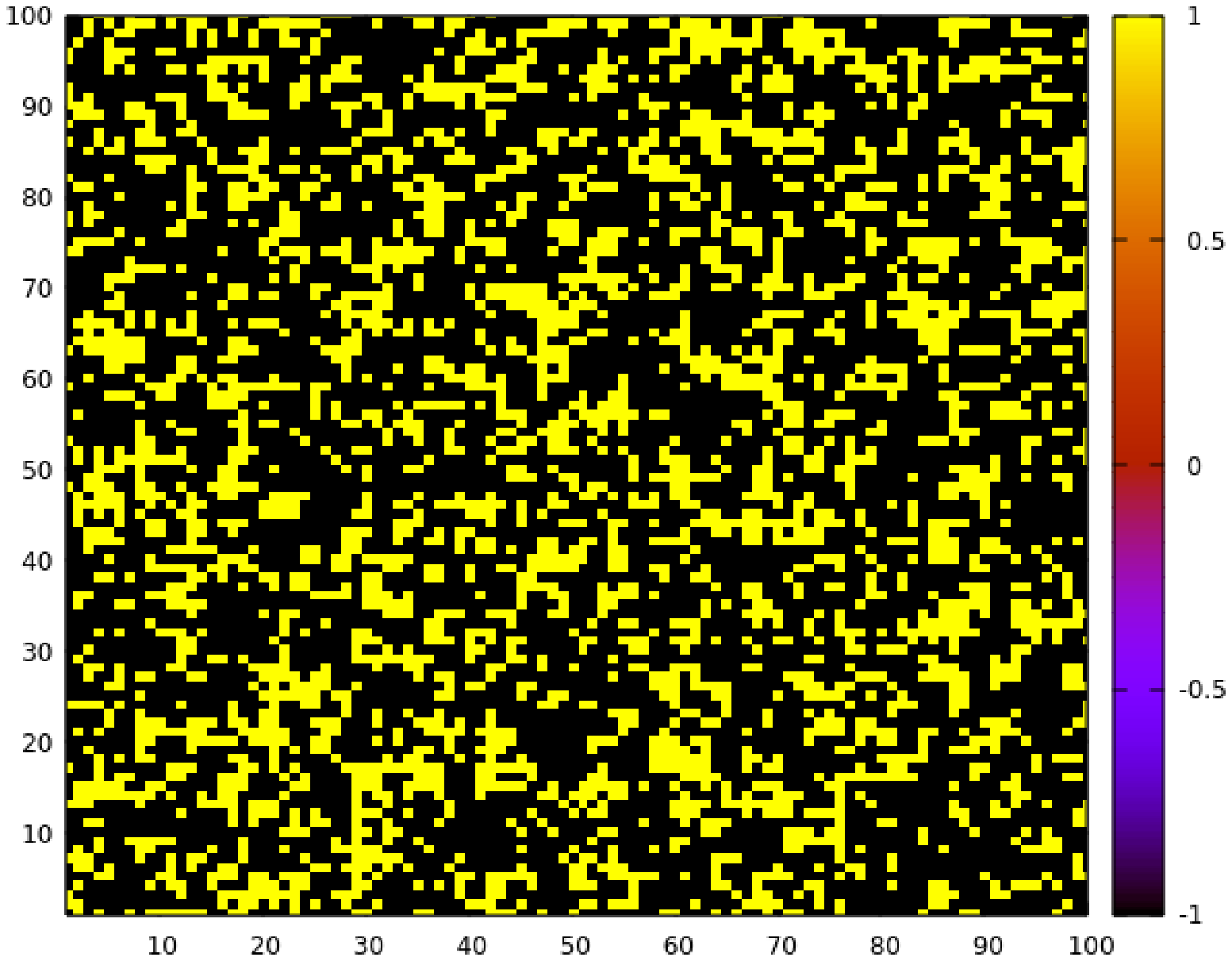}}
			(b)
			\resizebox{5.0cm}{!}{\includegraphics[angle=0]{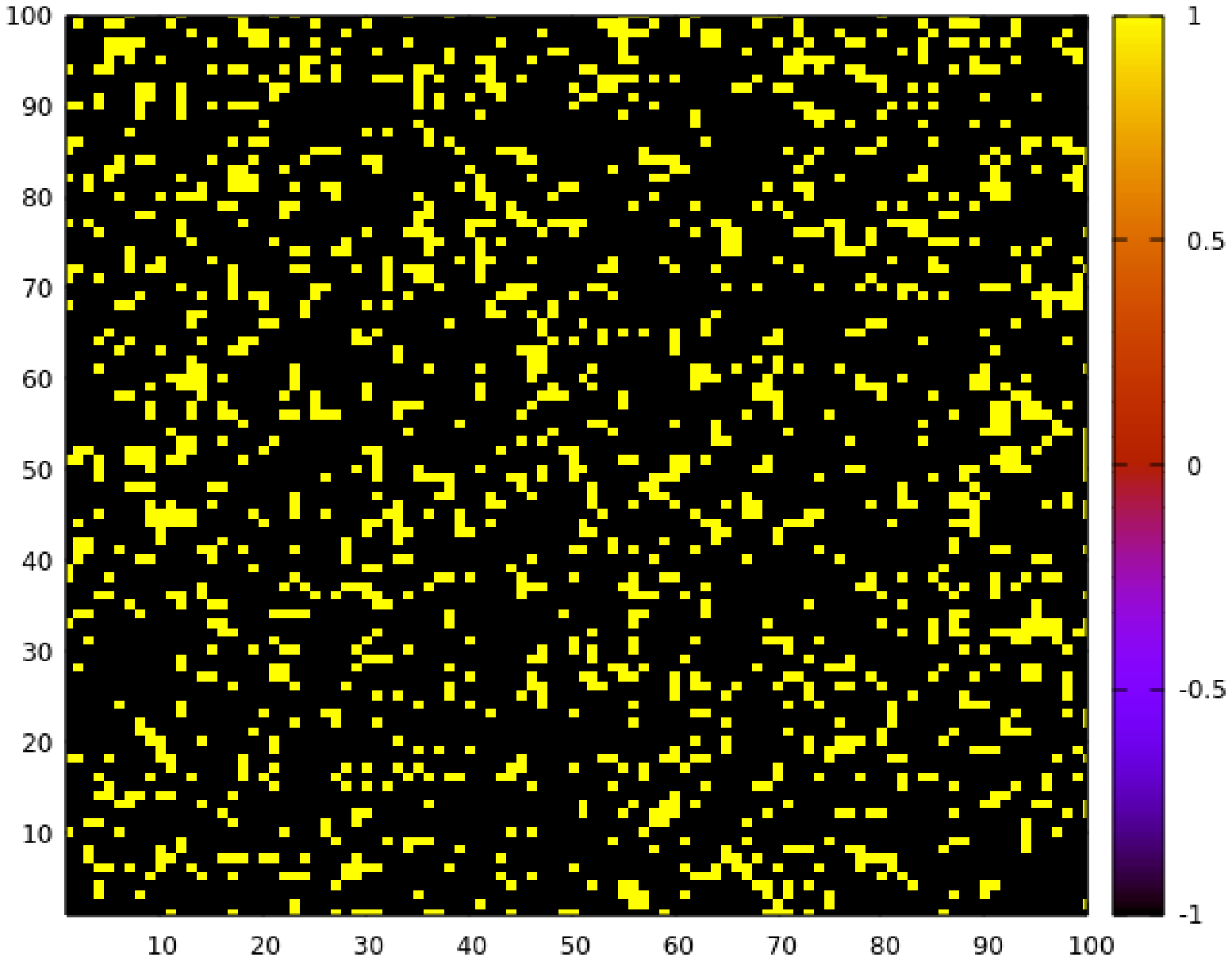}}
			(c)
			\resizebox{5.0cm}{!}{\includegraphics[angle=0]{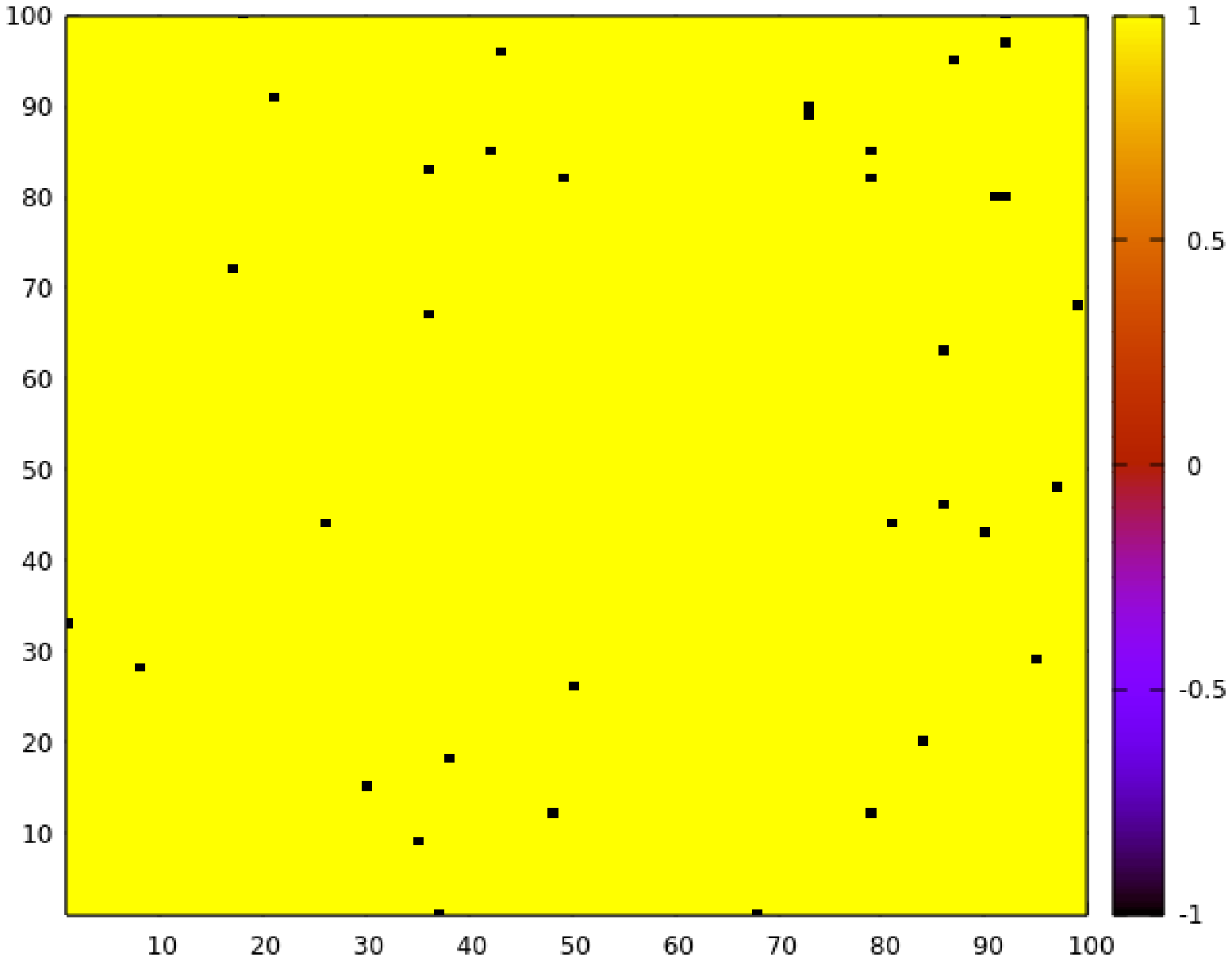}}
		\end{tabular}
		\caption{Morphology of (a) Top layer, (b) Mid layer and (c) Bottom layer for the AAB stacking ($J_{AA}/J_{BB}=0.3$ and $J_{AB}/J_{BB}=-0.4$) at $T=1.4$ (immediate lower neighbour than $T_{comp}$) with $M_{t}=-0.39$, $M_{m}=-0.69$, $M_{b}=0.99$.}		
		\label{fig_aab_morpho3}
	\end{center}
\end{figure}

\begin{center} {\large \textbf {a. Morphology}}\end{center}
The density plots of the spin matrices of three layers, in the presence of compensation effect for both the configurations, show that the critical temperature and the compensation temperature are morphologically different.\\
At $T_{comp}$, due to the configuration of AAB system,  compensation happens for satisfying the following conditions (which can readily be verified from Fig.-\ref{fig_aab_morpho1}, Fig.-\ref{fig_aab_morpho2} \& Fig.-\ref{fig_aab_morpho3}):
\begin{eqnarray}
\lvert M_{b}\lvert=\lvert M_{t}+M_{m}\lvert\\
sgn(M_{b})=-sgn(M_{t})\\
sgn(M_{b})=-sgn(M_{m})
\end{eqnarray}
Fig.-\ref{fig_aab_morpho1}, Fig.-\ref{fig_aab_morpho2} and Fig.-\ref{fig_aab_morpho3} are spin density matrix plots for AAB system with $J_{AA}/J_{BB}=0.3$ and $J_{AB}/J_{BB}=-0.4$. For such a configuration, the bottom-layer, $b$, is antiferromagnetically coupled to the mid-layer, $m$ and thus magnetically saturates in the opposite direction to both, $t$ and $m$. At $T_{crit}$, larger spin clusters results in bottom layer having greater absolute value of magnetisation than the top and mid layer.

\begin{figure}[!htb]
	\begin{center}
		\begin{tabular}{c}
				(a)
				\resizebox{5.0cm}{!}{\includegraphics[angle=0]{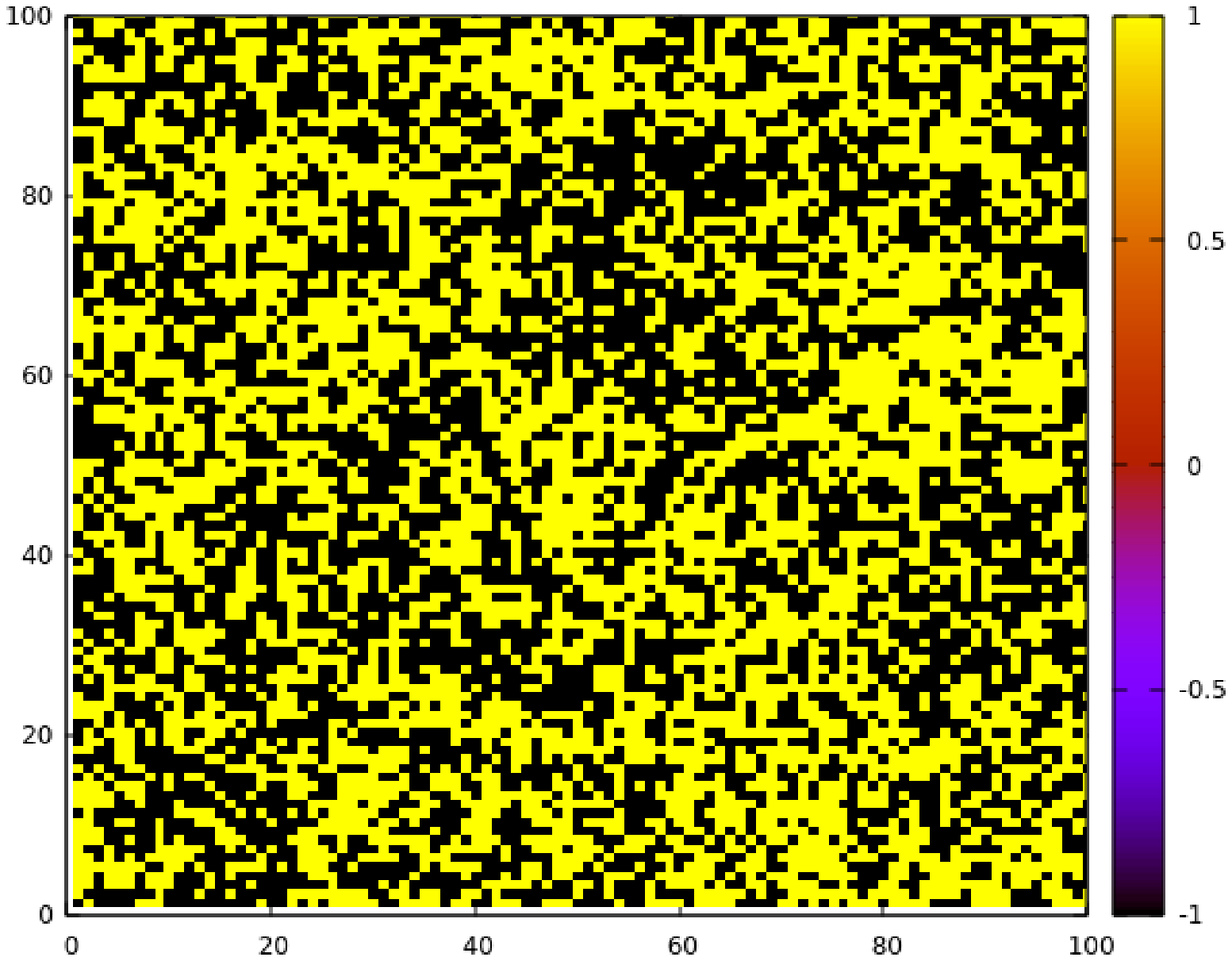}}
				(b)
				\resizebox{5.0cm}{!}{\includegraphics[angle=0]{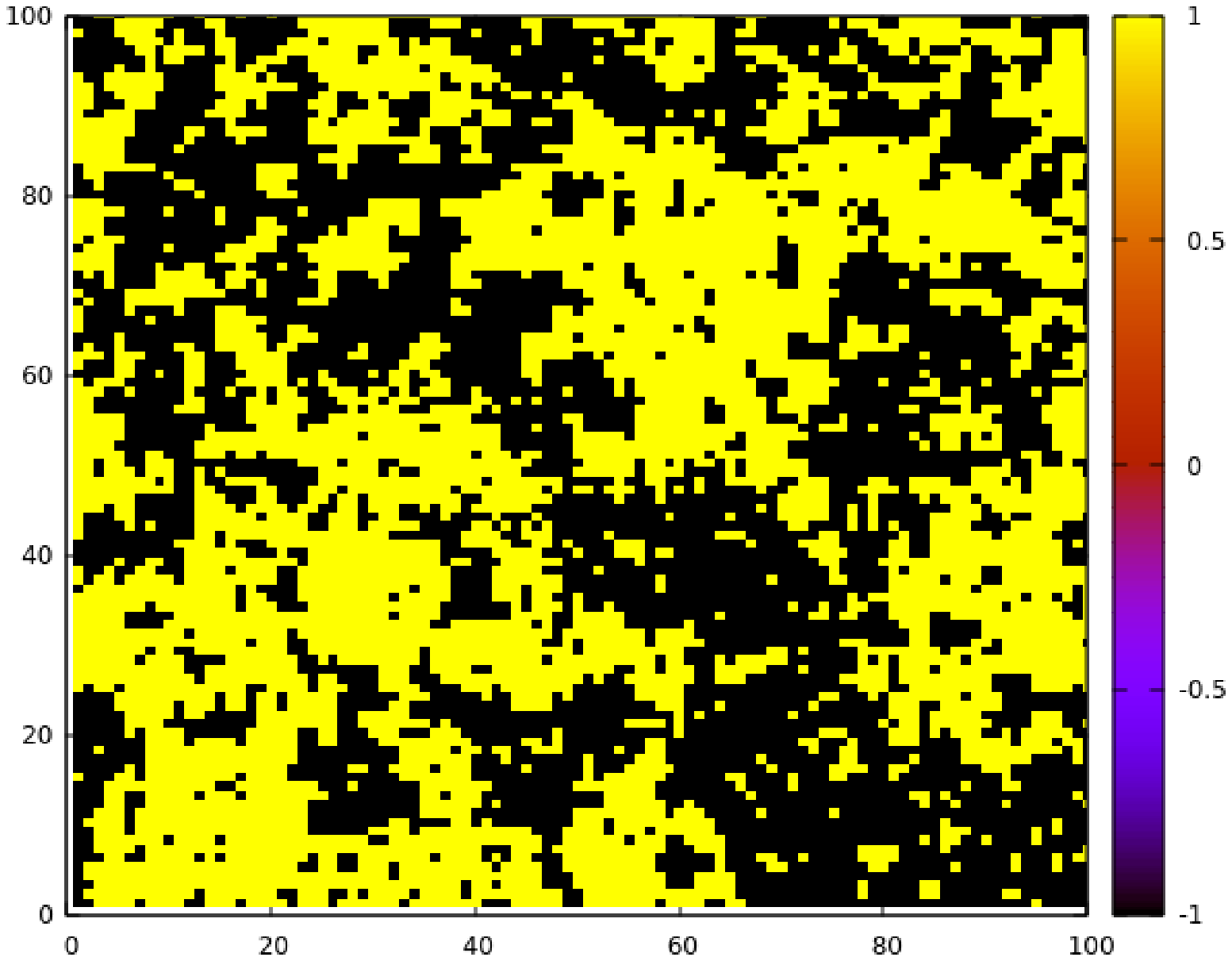}}
				(c)
				\resizebox{5.0cm}{!}{\includegraphics[angle=0]{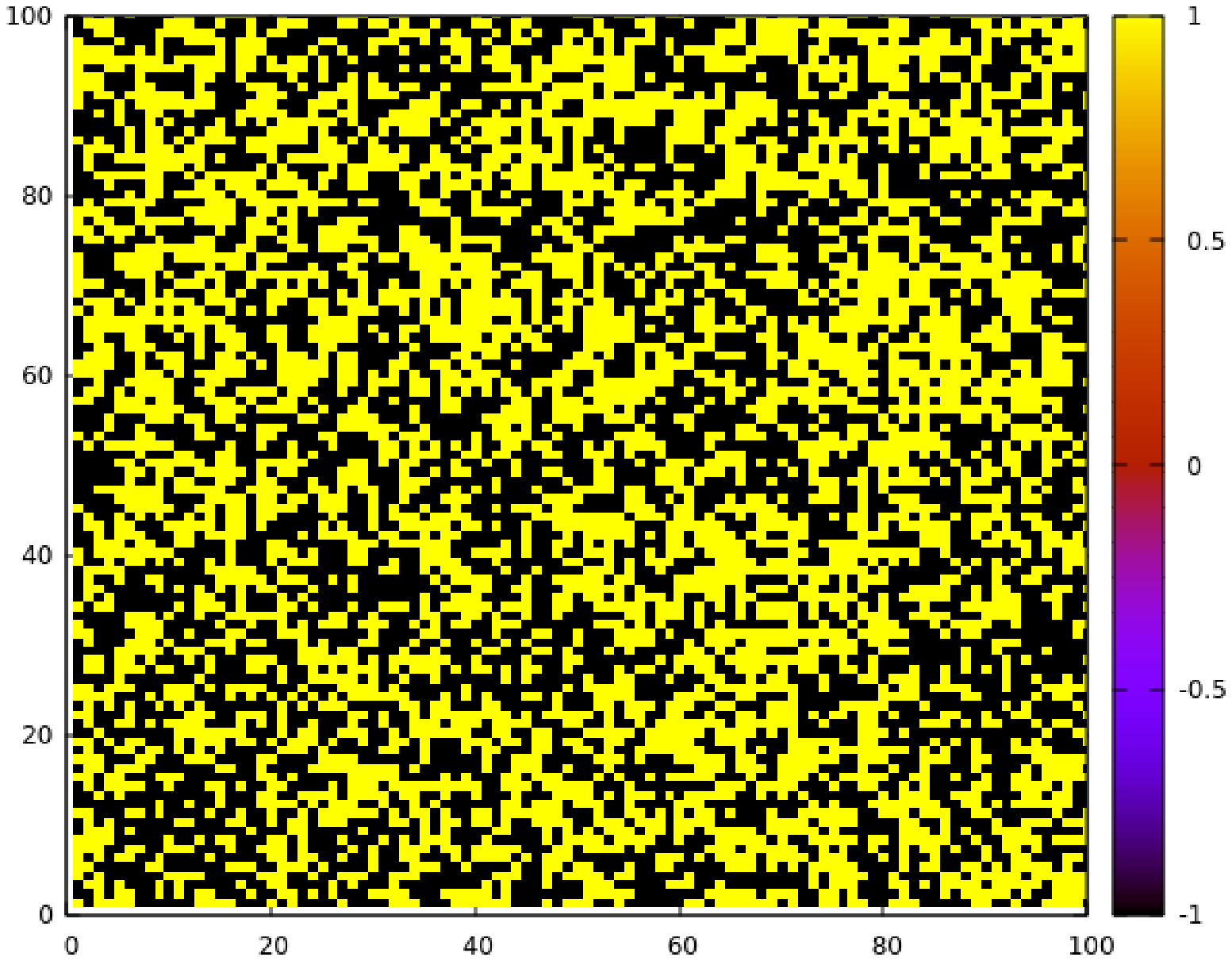}}
			\end{tabular}
			\caption{Morphology of (a) Top layer, (b) Mid layer and (c) Bottom layer for the ABA stacking ($J_{AA}/J_{BB}=0.3$ and $J_{AB}/J_{BB}=-0.4$) at $T_{crit}=2.6$ with $M_{t}=-2.91\times10^{-3}$, $M_{m}=1.23\times10^{-2}$, $M_{b}=-3.07\times10^{-3}$.}		
			\label{fig_aba_morpho1}
	\end{center}
\end{figure}

\begin{figure}[!htb]
	\begin{center}
		\begin{tabular}{c}
			(a)
			\resizebox{5.0cm}{!}{\includegraphics[angle=0]{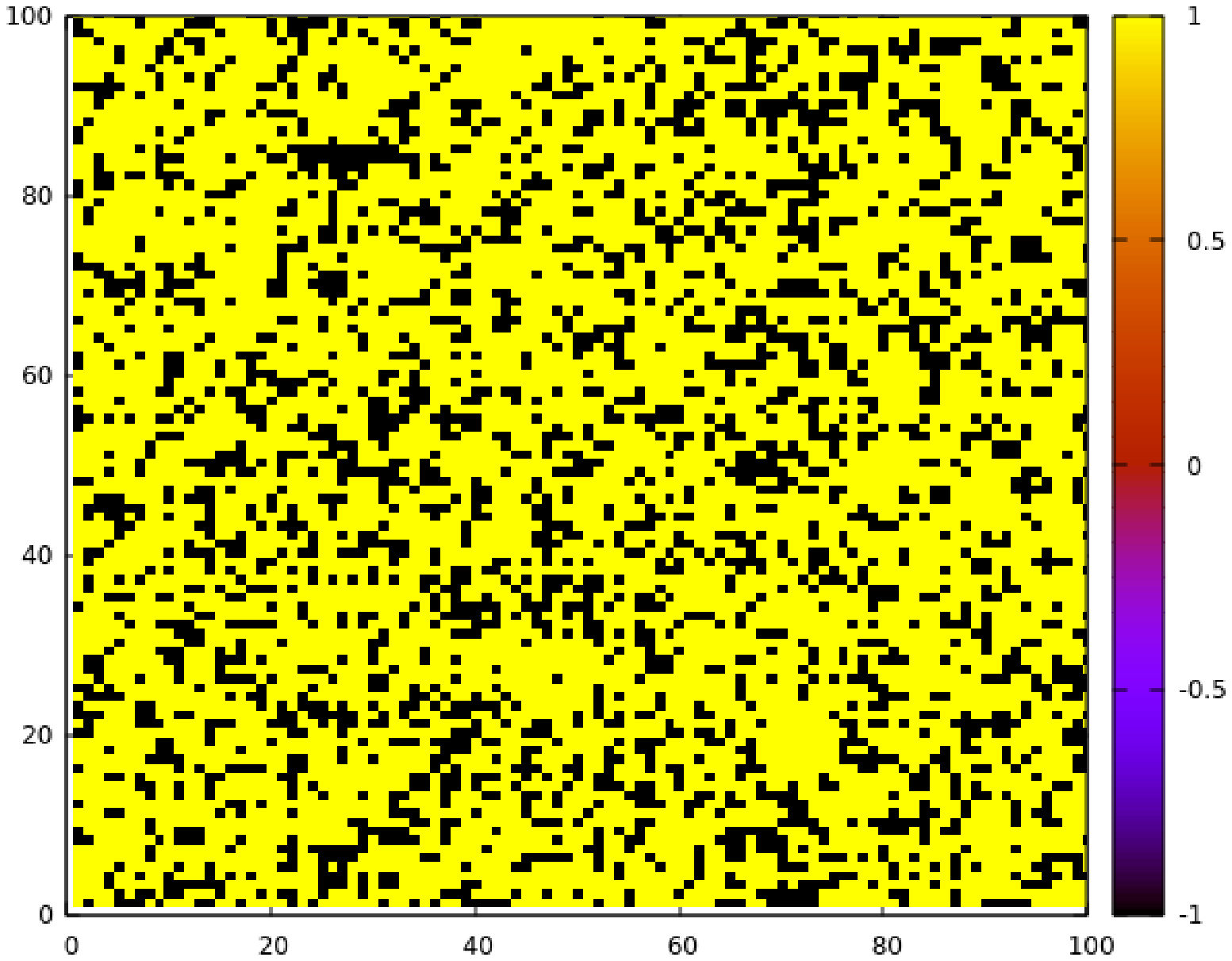}}
			(b)
			\resizebox{5.0cm}{!}{\includegraphics[angle=0]{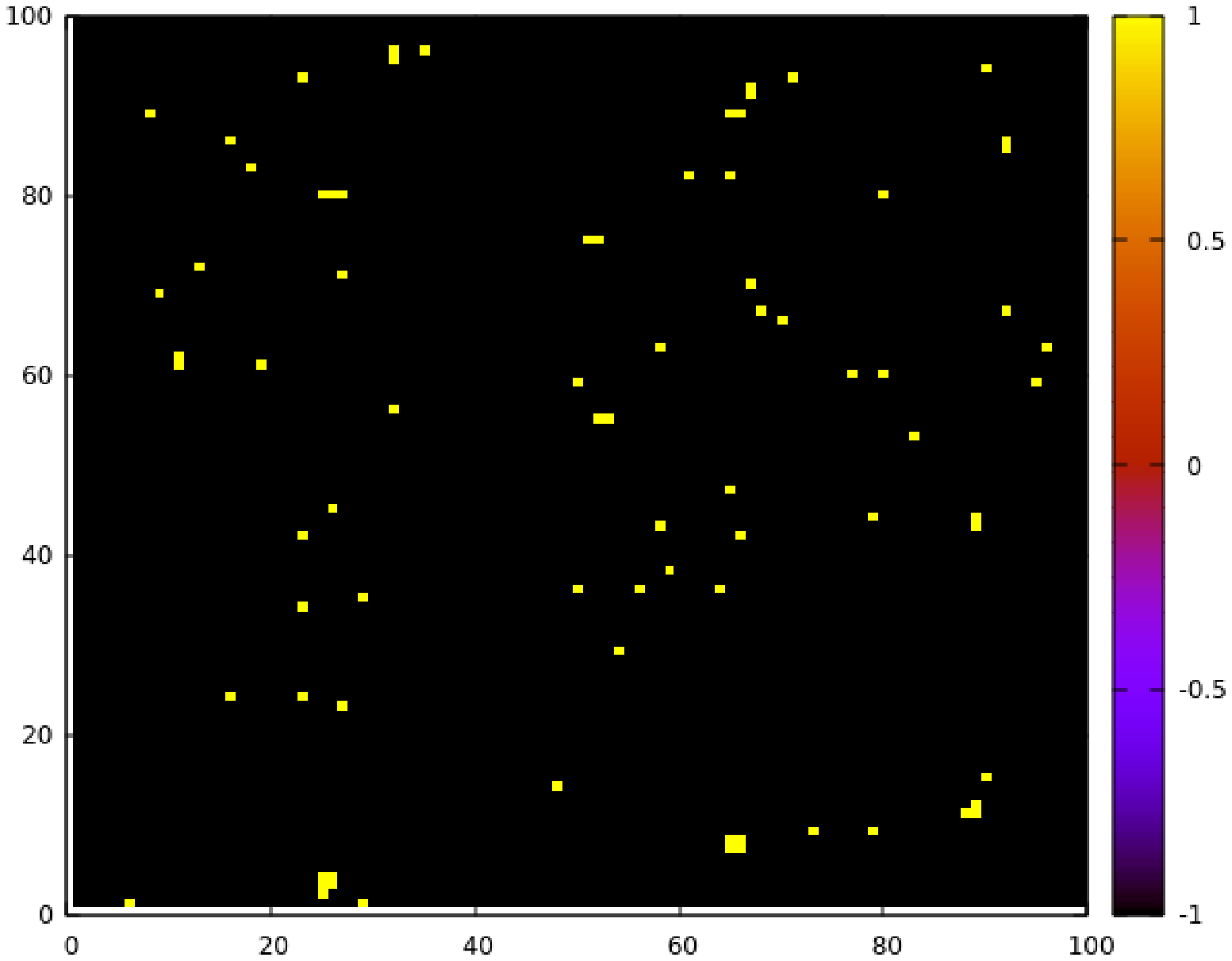}}
			(c)
			\resizebox{5.0cm}{!}{\includegraphics[angle=0]{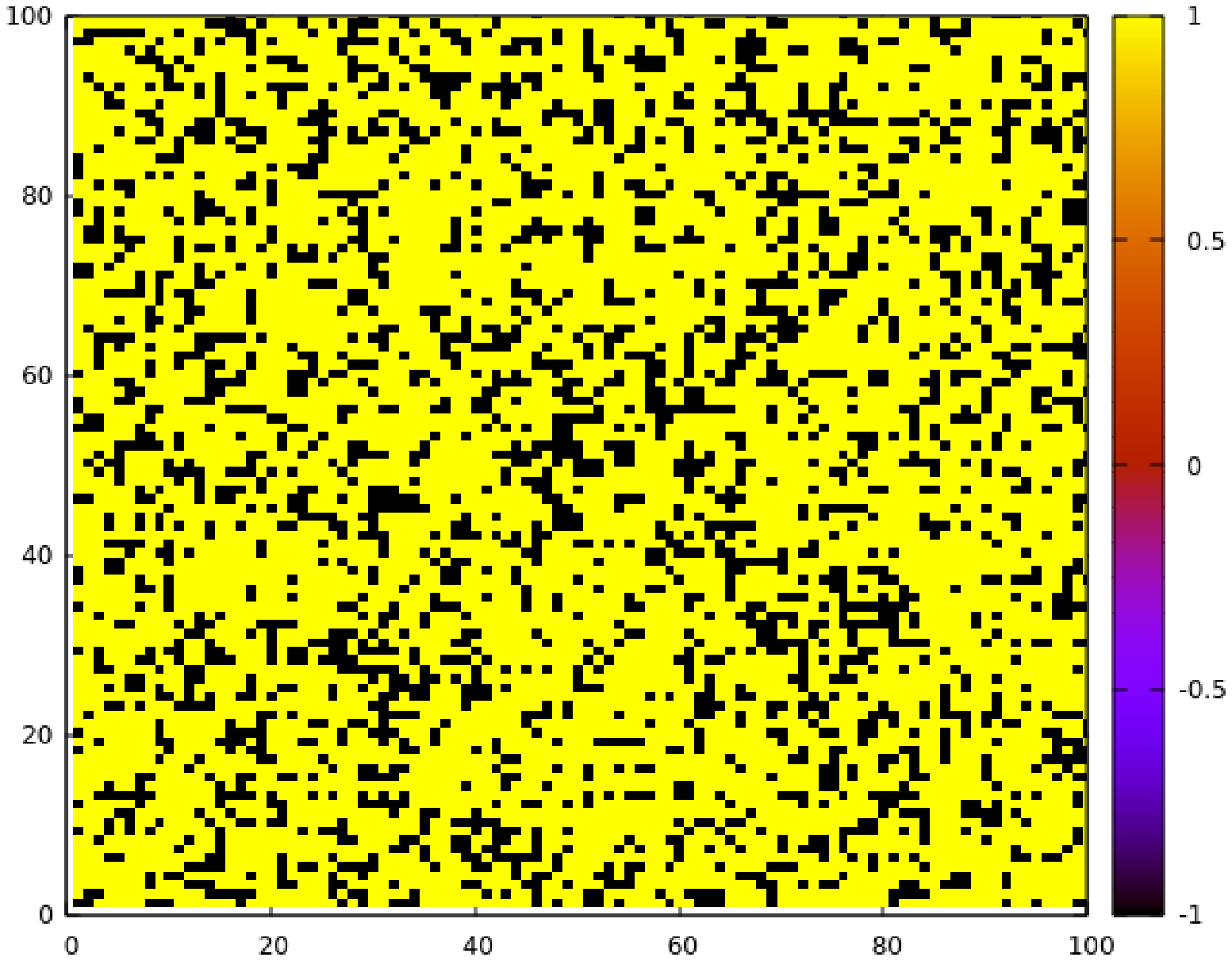}}
		\end{tabular}
		\caption{Morphology of (a) Top layer, (b) Mid layer and (c) Bottom layer for the ABA stacking ($J_{AA}/J_{BB}=0.3$ and $J_{AB}/J_{BB}=-0.4$) at $T=1.7$, (immediate higher neighbour than $T_{comp}$) with $M_{t}=-0.48$, $M_{m}=0.98$, $M_{b}=-0.48$.}		
		\label{fig_aba_morpho2}
	\end{center}
\end{figure}

\begin{figure}[!htb]
	\begin{center}
		\begin{tabular}{c}
			(a)
			\resizebox{5.0cm}{!}{\includegraphics[angle=0]{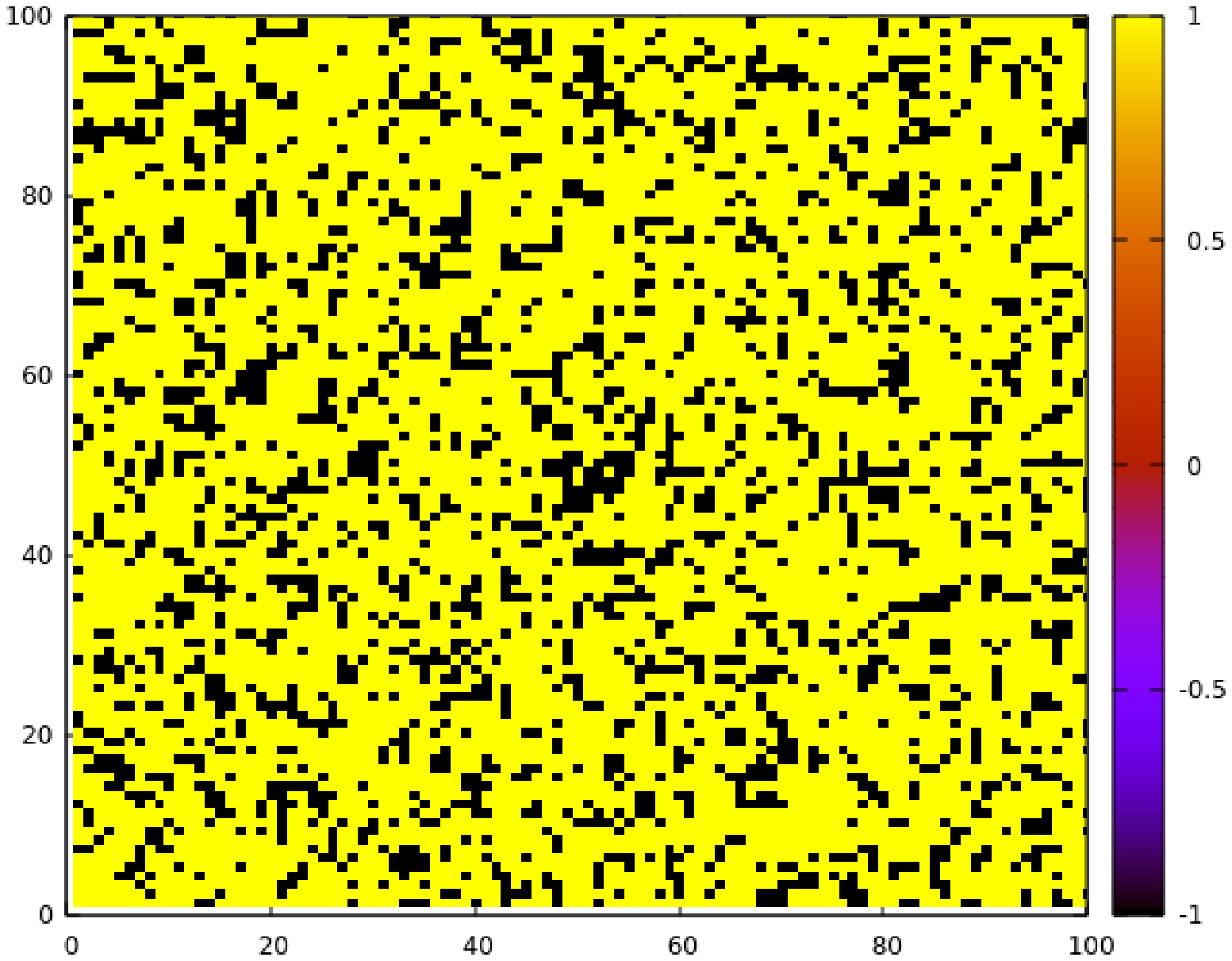}}
			(b)
			\resizebox{5.0cm}{!}{\includegraphics[angle=0]{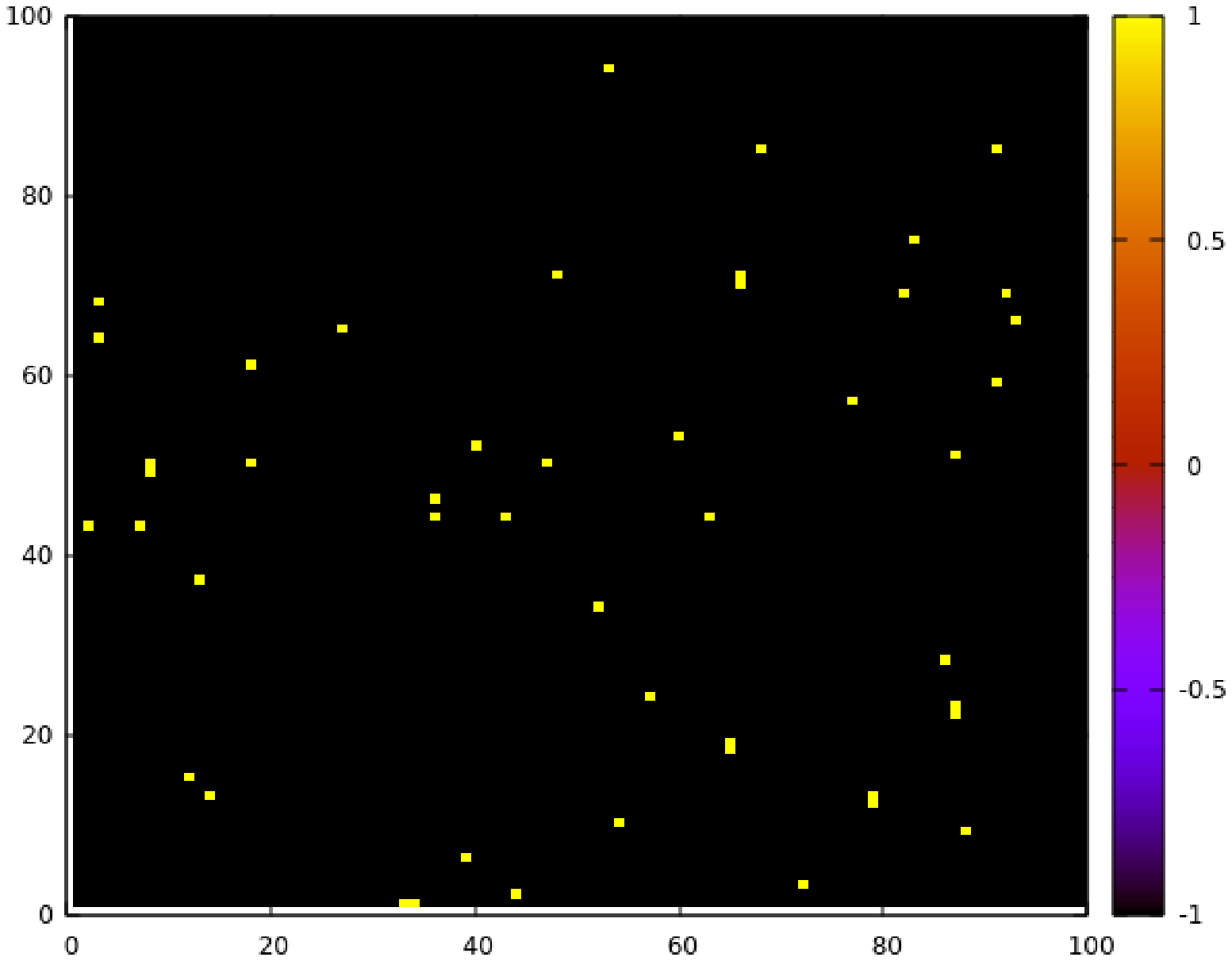}}
			(c)
			\resizebox{5.0cm}{!}{\includegraphics[angle=0]{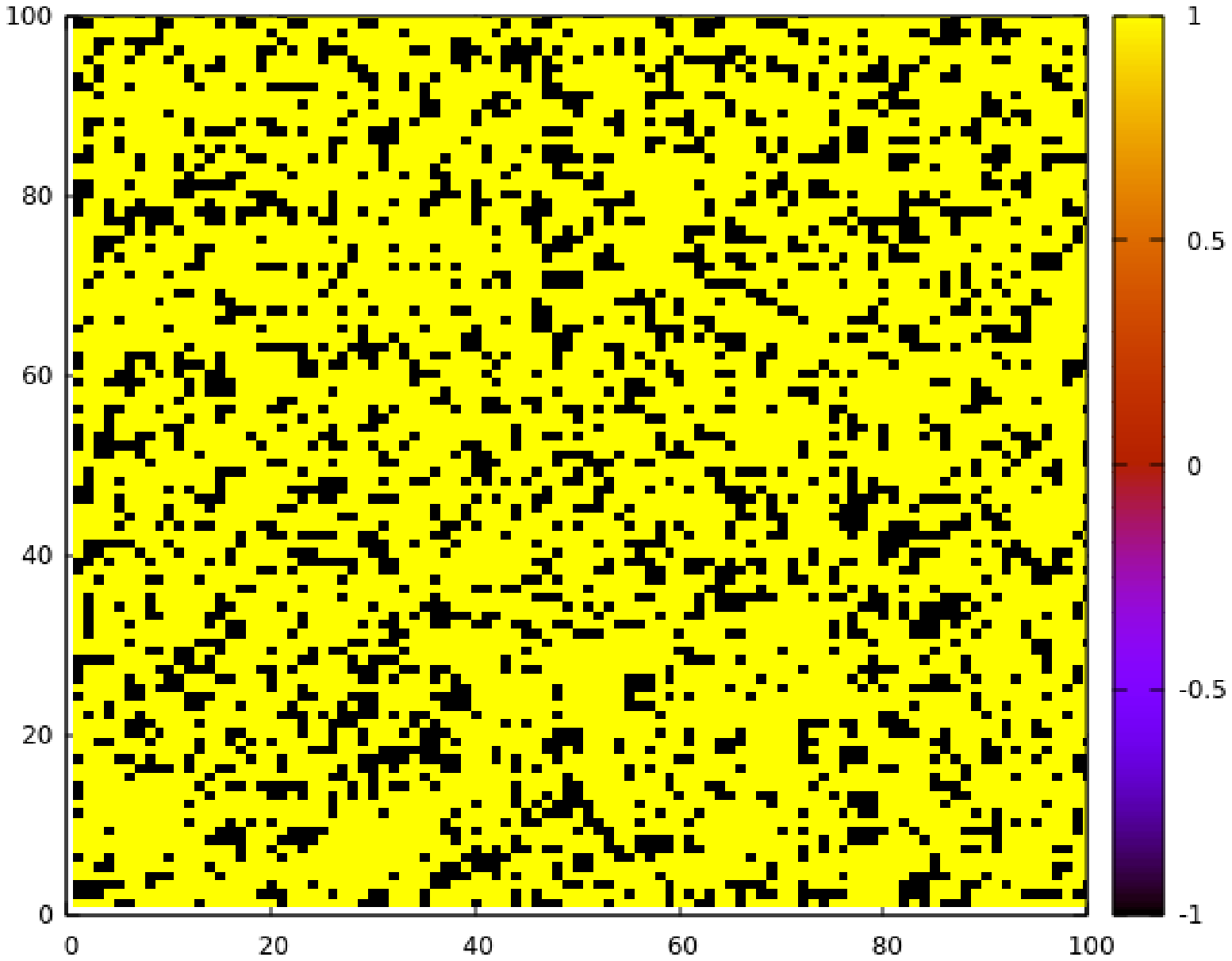}}
		\end{tabular}
		\caption{Morphology of (a) Top layer, (b) Mid layer and (c) Bottom layer for the ABA stacking ($J_{AA}/J_{BB}=0.3$ and $J_{AB}/J_{BB}=-0.4$) at $T=1.6$ (immediate lower neighbour than $T_{comp}$) with $M_{t}=-0.53$, $M_{m}=0.99$, $M_{b}=-0.53$.}		
		\label{fig_aba_morpho3}
	\end{center}
\end{figure}

\noindent Similarly for ABA type systems, the conditions for compensation change to the following:
\begin{eqnarray}
\lvert M_{m}\lvert=\lvert M_{t}+M_{b}\lvert\\
sgn(M_{m})=-sgn(M_{t})\\
sgn(M_{m})=-sgn(M_{b})
\end{eqnarray}
Fig.-\ref{fig_aba_morpho1}, Fig.-\ref{fig_aba_morpho2}, Fig.-\ref{fig_aba_morpho3} are spin density matrix plots for ABA system with $J_{AA}/J_{BB}=0.3$ and $J_{AB}/J_{BB}=-0.4$.
As the mid-layer,$m$, is antiferromagnetically coupled to both the $t$ and $b$ layers, mid-layer always magnetically saturates in the opposite direction to both, $t$ and $b$ layers for an ABA system. At $T_{crit}$, it is seen that larger spin clusters form in the mid layer resulting in its having greater absolute value of magnetisation than the top and bottom layer.

\begin{figure}[!htb]
	\begin{center}
		\begin{tabular}{c}
			(a)
			\resizebox{9.5cm}{!}{\includegraphics[angle=0]{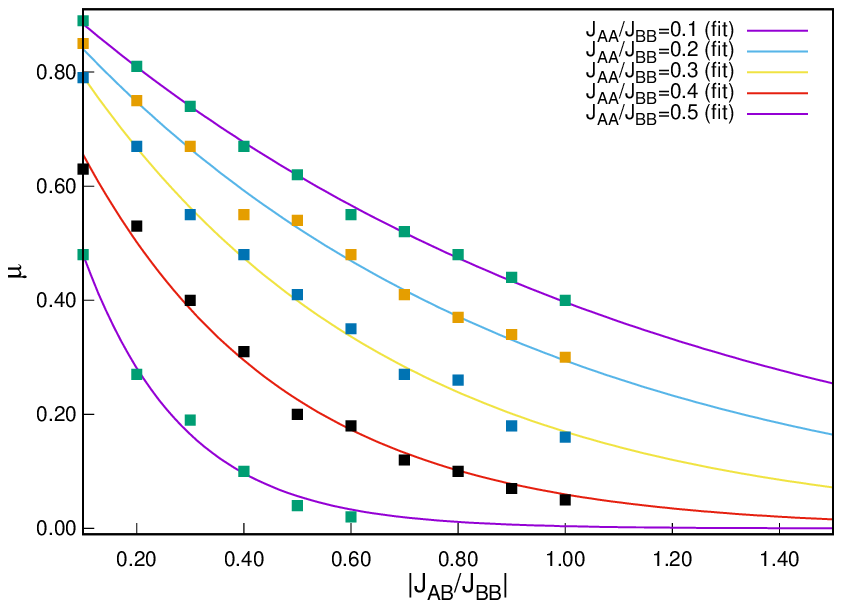}}
			(b)
			\resizebox{9.5cm}{!}{\includegraphics[angle=0]{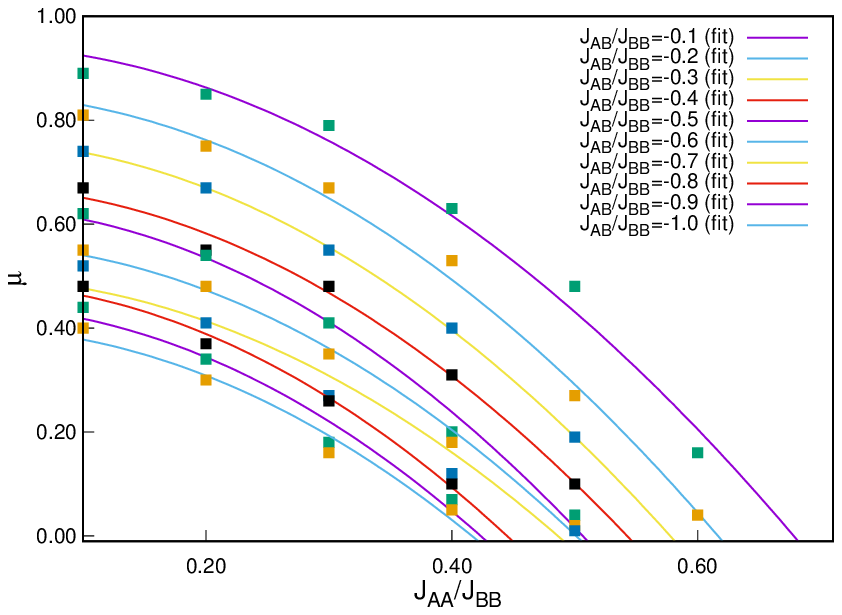}}
		\end{tabular}
		\caption{Plots of (a) $\Phi_{1}$ vs. $\left|\dfrac{J_{AB}}{J_{BB}}\right|$ (b) $\Phi_{2}$ vs. $\dfrac{J_{AA}}{J_{BB}}$ to show probable dependance of IARRM of an AAB configuration.}		
		\label{fig_phi1_phi2}
	\end{center}
\end{figure}

\begin{center} {\large \textbf {b. AAB composition}}\end{center}
{\bf Inverse Absolute of Reduced Residual Magnetisation (IARRM):}
\vspace{5pt}
\\ The behaviour of IARRM with the variation of antiferromagnetic  and ferromagnetic coupling strength ratios (AFM and FM ratios, for brevity) [Fig.-\ref{fig_aab_rrm_3d}] leads us to predict their behaviours in the following way:
\begin{figure}[!htb]
	\begin{center}
		\begin{tabular}{c}
			(a)
			\resizebox{8.5cm}{!}{\includegraphics[angle=0]{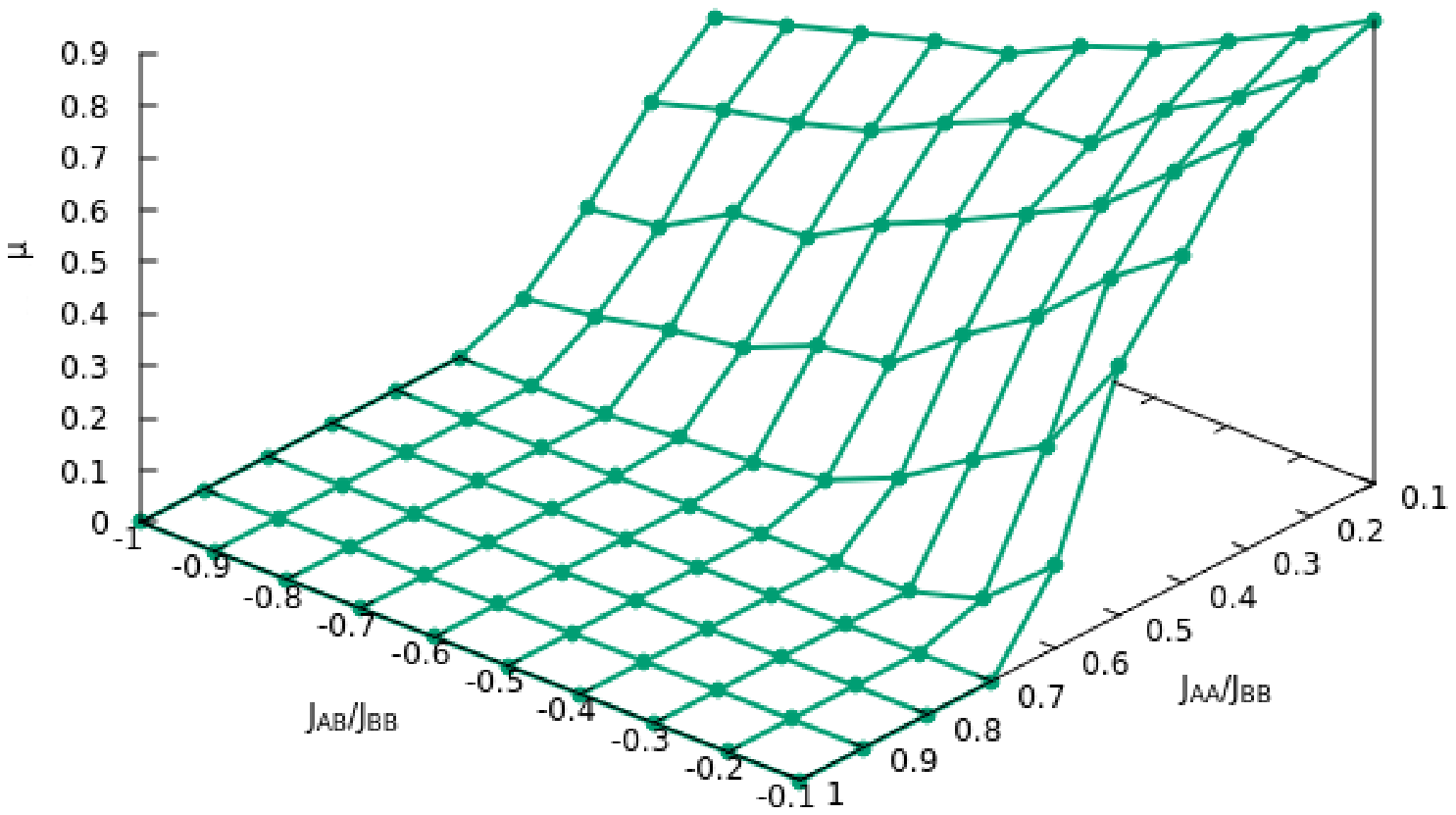}}
			(b)
			\resizebox{8.5cm}{!}{\includegraphics[angle=0]{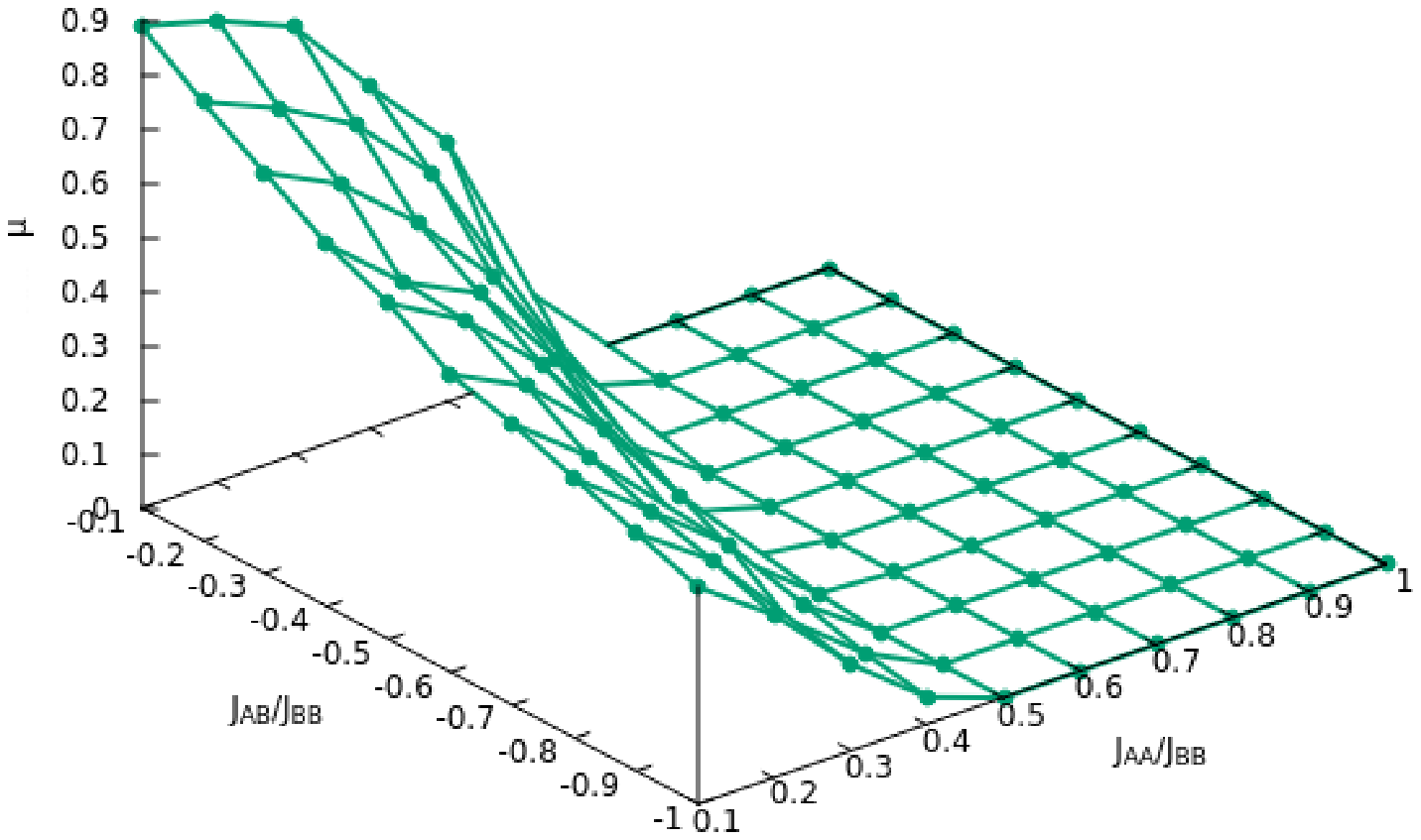}}
		\end{tabular}
		\caption{3D plots of IARRM vs. controlling factors for AAB configuration from two different angles of view.}		
		\label{fig_aab_rrm_3d}
	\end{center}
\end{figure}
The plots of IARRM vs. AFM ratio [Fig.-\ref{fig_phi1_phi2}a], \textit{for a fixed FM ratio}, predicts a functional relationship, $\Phi_{1}\left(J_{AA}/J_{BB}, J_{AB}/J_{BB}\right)$, like:
\begin{equation}
\Phi_{1}\left(J_{AA}/J_{BB}, J_{AB}/J_{BB}\right) |=a_{1}\exp{\left[-a_{2}\left|\dfrac{J_{AB}}{J_{BB}}\right|\right]}
\end{equation}
where the coefficients $a_{1}$ and $a_{2}$ are functions of FM ratios i.e. $a_{1}\equiv a_{1}(J_{AA}/J_{BB}) \text{ and } a_{2}\equiv a_{2}(J_{AA}/J_{BB})$ and Appendix, A1: Table \ref{tab_a1_a2} contains their variations.
\\\indent The plots of IARRM vs. FM ratio [Fig.-\ref{fig_phi1_phi2}b], \textit{for fixed AFM ratios}, predict a functional dependence, $\Phi_{2}\left(J_{AA}/J_{BB}, J_{AB}/J_{BB}\right)$, like:
\begin{equation}
\Phi_{2}\left(J_{AA}/J_{BB}, J_{AB}/J_{BB}\right)= a_{3}-a_{4}\left(\dfrac{J_{AA}}{J_{BB}}\right)^2
\end{equation}
where the coefficients $a_{3}$ and $a_{4}$ are functions of $J_{AB}/J_{BB}$ i.e. $a_{3}\equiv a_{3}(J_{AB}/J_{BB})$ and $a_{4}\equiv a_{4}(J_{AB}/J_{BB})$. Variations of $a_{3}$ and $a_{4}$ are listed in Appendix, A1: Table \ref{tab_a3_a4}.

\begin{figure}[!htb]
	\begin{center}
		\begin{tabular}{c}
			(a)
			\resizebox{8.5cm}{!}{\includegraphics[angle=0]{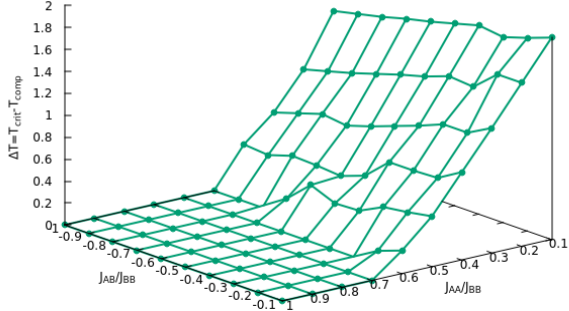}}
			(b)
			\resizebox{8.5cm}{!}{\includegraphics[angle=0]{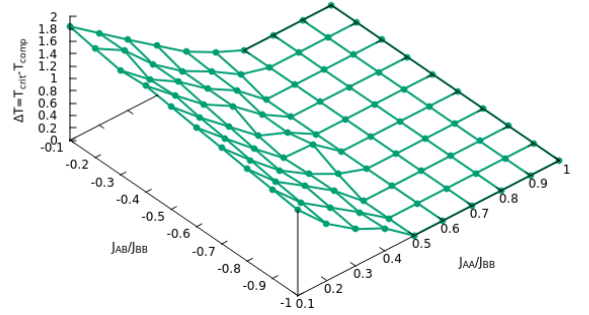}}
		\end{tabular}
		\caption{3D plots of $\Delta T=T_{crit}-T_{comp}$ vs. controlling factors for AAB configuration from two different angles of view.}		
		\label{fig_aab_critcomp_3d}
	\end{center}
\end{figure}

\begin{figure}[!htb]
	\begin{center}
		\begin{tabular}{c}
			(a)
			\resizebox{9.5cm}{!}{\includegraphics[angle=0]{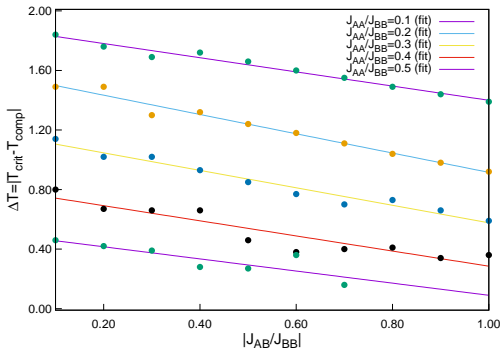}}
			(b)
			\resizebox{9.5cm}{!}{\includegraphics[angle=0]{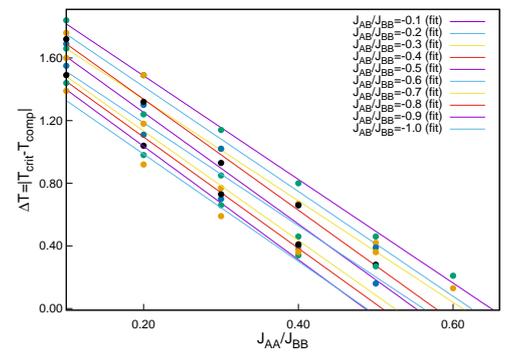}}
		\end{tabular}
		\caption{Plots of (a) $\Psi_{1}$ vs. $\left|\dfrac{J_{AB}}{J_{BB}}\right|$ (b) $\Psi_{2}$ vs. $\dfrac{J_{AA}}{J_{BB}}$ to show probable dependence of $\Delta T$ of an AAB configuration.}		
		\label{fig_aab_critcomp}
	\end{center}
\end{figure}

\vspace{15pt}
{\bf \noindent Temperature gap between Critical and Compensation temperatures, $\Delta T$ :}
\vspace{5pt}
\\The behaviour of temperature gap between Critical and Compensation temperatures with the variation of AFM and FM ratios [Fig.-\ref{fig_aab_critcomp_3d}], leads us to predict its behaviour in similar fashion as before.
\\The plots of $\Delta T$ vs. AFM ratio [Fig.-\ref{fig_aab_critcomp}a], \textit{for fixed FM ratios}, predict a functional relationship, $\Psi_{1}\left(J_{AA}/J_{BB}, J_{AB}/J_{BB}\right)$, like:
\begin{equation}
\Psi_{1}\left(J_{AA}/J_{BB}, J_{AB}/J_{BB}\right)=-a_{5}\left|\dfrac{J_{AB}}{J_{BB}}\right|+a_{6}
\end{equation}
where the coefficients $a_{5}$ and $a_{6}$ are functions of FM ratios i.e. $a_{5}\equiv a_{5}(J_{AA}/J_{BB})$ and $a_{6}\equiv a_{6}(J_{AA}/J_{BB})$. Changes in the values of $a_{5}$ and $a_{6}$ are listed in Appendix, A1: Table \ref{tab_a5_a6}.

\begin{figure}[!htb]
	\begin{center}
		\begin{tabular}{c}
			(a)
			\resizebox{8.5cm}{!}{\includegraphics[angle=0]{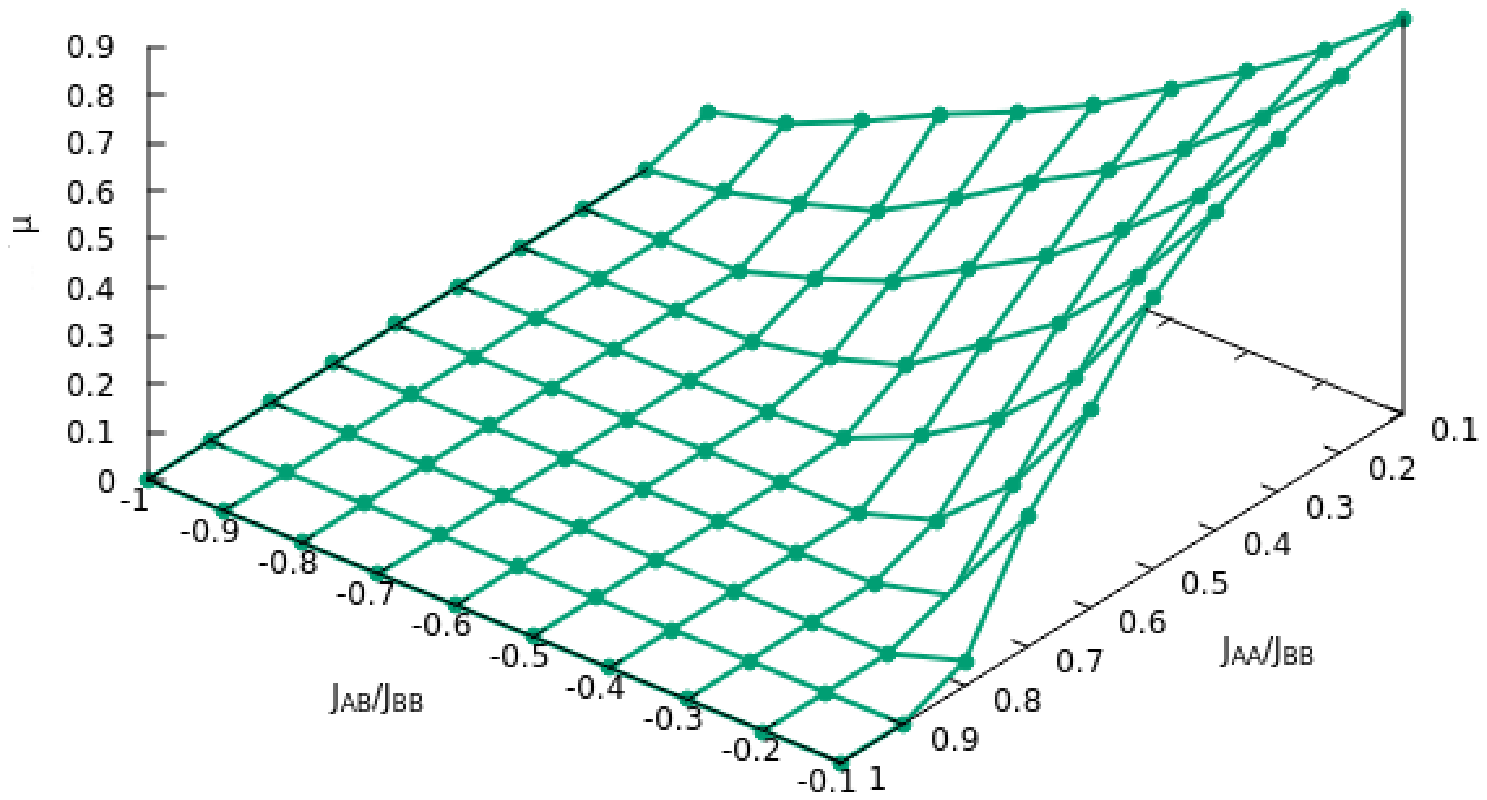}}
			(b)
			\resizebox{8.5cm}{!}{\includegraphics[angle=0]{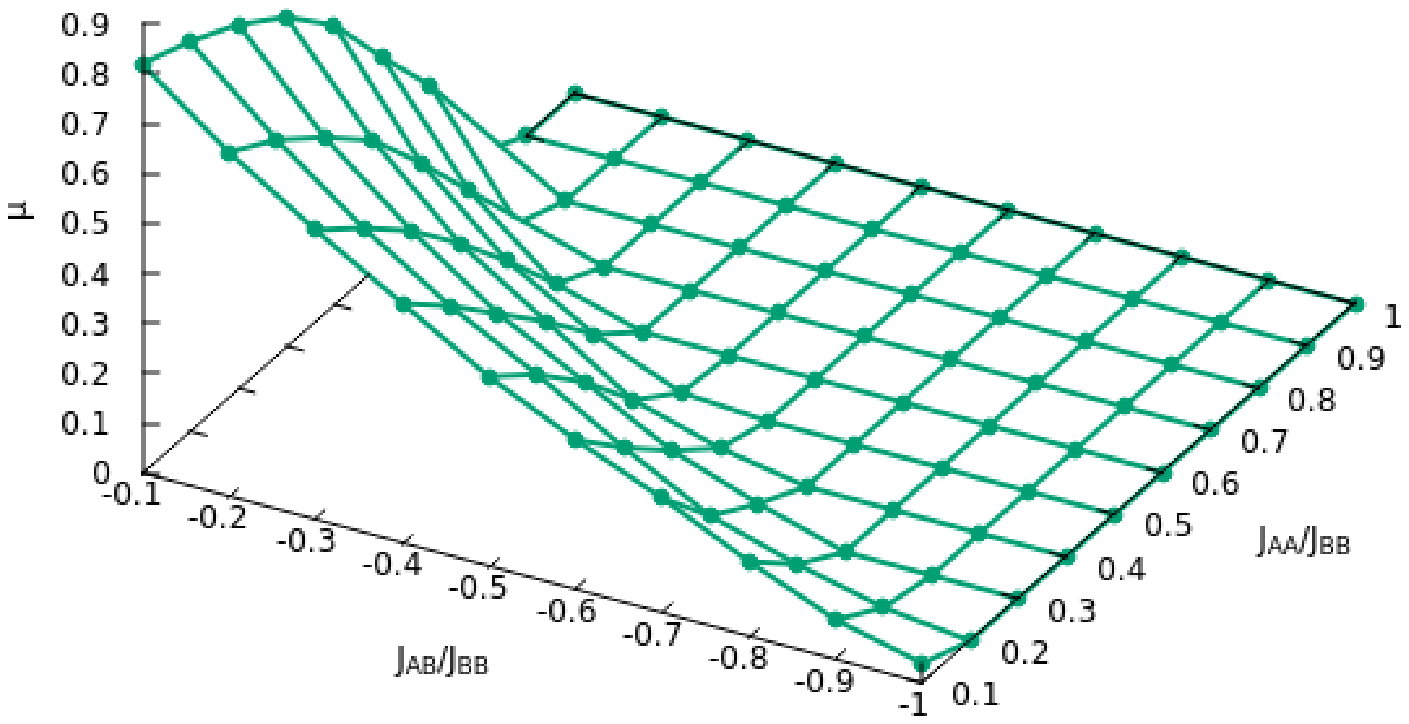}}
		\end{tabular}
		\caption{3D plots of IARRM vs. controlling factors for ABA configuration from two different angles of view.}		
		\label{fig_aba_rrm_3d}
	\end{center}
\end{figure}

\begin{figure}[!htb]
	\begin{center}
		\begin{tabular}{c}
			(a)
			\resizebox{9.5cm}{!}{\includegraphics[angle=0]{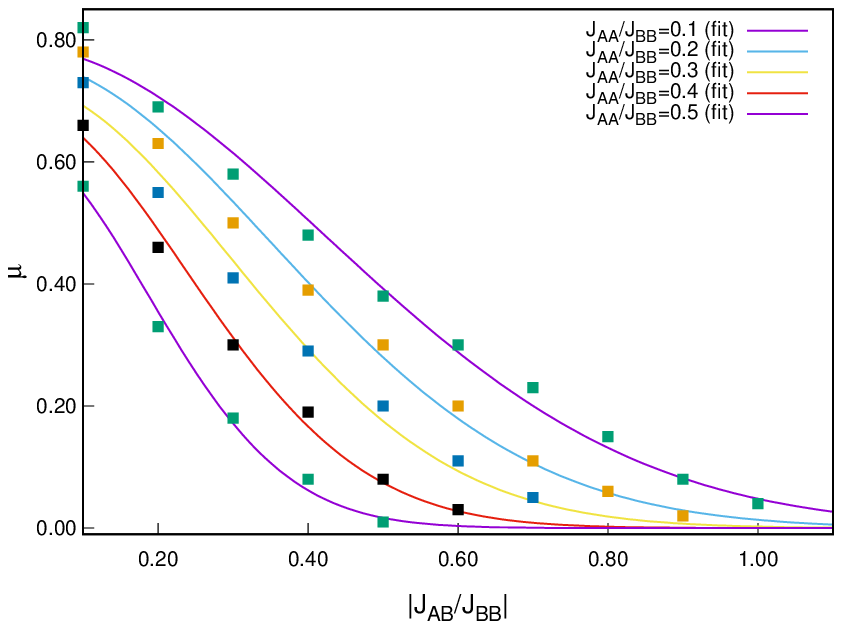}}
			(b)
			\resizebox{9.5cm}{!}{\includegraphics[angle=0]{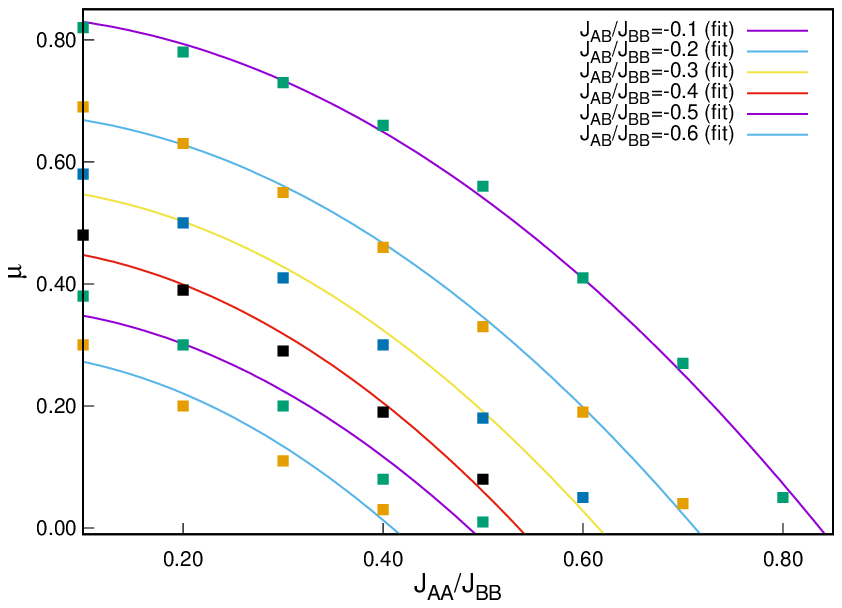}}
		\end{tabular}
		\caption{Plots of (a) $\Phi_{3}$ vs. $\left|\dfrac{J_{AB}}{J_{BB}}\right|$ (b) $\Phi_{4}$ vs. $\dfrac{J_{AA}}{J_{BB}}$ to show probable dependence of $\Delta T$ of an ABA configuration.}		
		\label{fig_aba_rrm}
	\end{center}
\end{figure}

\begin{figure}[!htb]
	\begin{center}
		\begin{tabular}{c}
			(a)
			\resizebox{9.5cm}{!}{\includegraphics[angle=0]{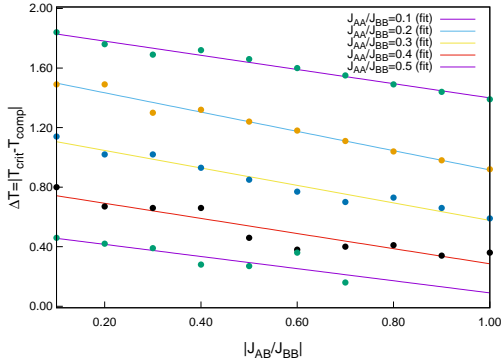}}
			(b)
			\resizebox{9.5cm}{!}{\includegraphics[angle=0]{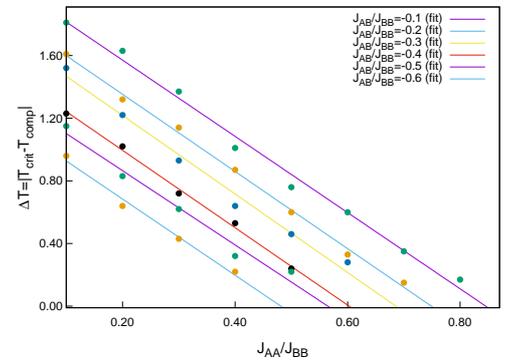}}
		\end{tabular}
		\caption{Plots of (a) $\Psi_{3}$ vs. $\left|\dfrac{J_{AB}}{J_{BB}}\right|$ (b) $\Psi_{4}$ vs. $\dfrac{J_{AA}}{J_{BB}}$ to show probable dependence of $\Delta T$ of an ABA configuration.}		
		\label{fig_aba_critcomp}
	\end{center}
\end{figure}

The plots of $\Delta T$ vs. FM ratio [Fig.-\ref{fig_aab_critcomp}b], \textit{for fixed AFM ratios}, predict a functional dependence, $\Psi_{2}\left(J_{AA}/J_{BB}, J_{AB}/J_{BB}\right)$, like:
\begin{equation}
\Psi_{2}\left(J_{AA}/J_{BB}, J_{AB}/J_{BB}\right)=-a_{7}\left(\dfrac{J_{AA}}{J_{BB}}\right)+a_{8}
\end{equation}
where the coefficients $a_{7}$ and $a_{8}$ are functions of $J_{AB}/J_{BB}$ i.e. $a_{7}\equiv a_{7}(J_{AB}/J_{BB})$ and $a_{8}\equiv a_{8}(J_{AB}/J_{BB})$. Variations of $a_{7}$ and $a_{8}$ are to be found in Appendix, A1: Table \ref{tab_a7_a8}.\\
\vspace{20pt}
\begin{center} {\large \textbf {c. ABA composition}}\end{center}
{\bf Inverse Absolute of Reduced Residual Magnetisation (IARRM):}
\vspace{5pt}
\\ Observing the trends from the plots of IARRM [Fig.-\ref{fig_aba_rrm_3d}, Fig.-\ref{fig_aba_rrm}a, Fig.-\ref{fig_aba_rrm}b], for the ABA configuration, like the AAB configuration, two functional relationships are predicted like, $\Phi_{3}\left(J_{AA}/J_{BB}, J_{AB}/J_{BB}\right)$ and $\Phi_{4}\left(J_{AA}/J_{BB}, J_{AB}/J_{BB}\right)$ as:
\begin{eqnarray}
\Phi_{3}\left(J_{AA}/J_{BB}, J_{AB}/J_{BB}\right) &=&b_{1}\exp{\left[-b_{2}\left(\dfrac{J_{AB}}{J_{BB}}\right)^2\right]} \\
\Phi_{4}\left(J_{AA}/J_{BB}, J_{AB}/J_{BB}\right)&=& b_{3}-b_{4}\left(\dfrac{J_{AA}}{J_{BB}}\right)^{2}
\end{eqnarray}
where the coefficients $b_{1}$ and $b_{2}$ are functions of FM ratios i.e. $b_{1}\equiv b_{1}(J_{AA}/J_{BB})$ and $b_{2}\equiv b_{2}(J_{AA}/J_{BB})$ and their variations are in Appendix, A3: Table \ref{tab_b1_b2} and the coefficients $b_{3}$ and $b_{4}$ are functions of AFM ratios i.e. $b_{3}\equiv b_{3}(J_{AB}/J_{BB})$ and $b_{4}\equiv b_{4}(J_{AB}/J_{BB})$ and their variations are in Appendix, A3: Table \ref{tab_b3_b4}. 
\vspace{15pt}
\\\noindent
{\bf Temperature gap between Critical and Compensation temperatures, $\Delta T$ :}
\vspace{5pt}
\\The plots of $\Delta T$ [Fig.-\ref{fig_aba_critcomp_3d}, Fig.-\ref{fig_aba_critcomp}a, Fig.-\ref{fig_aba_critcomp}b] may lead to functional relationships like:
\begin{figure}[!htb]
	\begin{center}
		\begin{tabular}{c}
			(a)
			\resizebox{8.5cm}{!}{\includegraphics[angle=0]{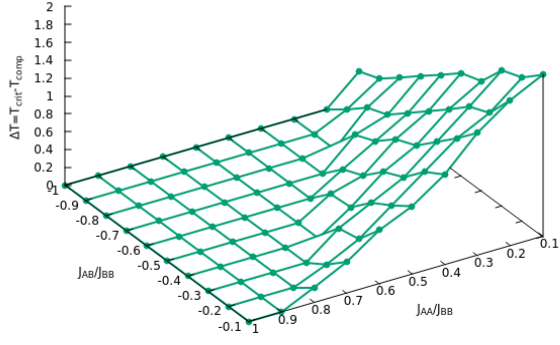}}
			(b)
			\resizebox{8.5cm}{!}{\includegraphics[angle=0]{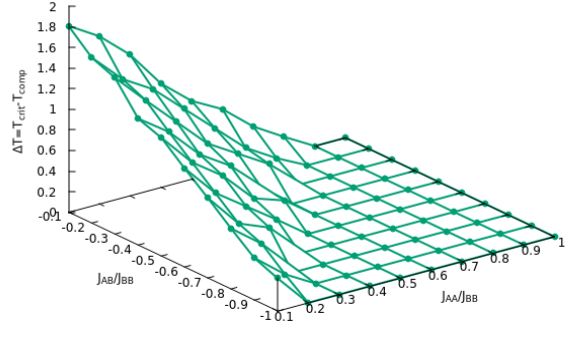}}
		\end{tabular}
		\caption{3D plots of $\Delta T$ vs. controlling factors for ABA configuration from two different angles of view.}		
		\label{fig_aba_critcomp_3d}
	\end{center}
\end{figure}
\begin{eqnarray}
\Psi_{3}\left(J_{AA}/J_{BB}, J_{AB}/J_{BB}\right)&=&-b_{5}\left|\dfrac{J_{AB}}{J_{BB}}\right|+b_{6}\\
\Psi_{4}\left(J_{AA}/J_{BB}, J_{AB}/J_{BB}\right)&=&-b_{7}\left(\dfrac{J_{AA}}{J_{BB}}\right)+b_{8}
\end{eqnarray}
where $b_{5}\equiv b_{5}(J_{AA}/J_{BB})$; $b_{6}\equiv b_{6}(J_{AA}/J_{BB})$ and $b_{7}\equiv b_{7}(J_{AB}/J_{BB})$; $b_{8}\equiv b_{8}(J_{AB}/J_{BB})$.\\
The obtained data of $b_{5}$ and $b_{6}$, are in Appendix, A3: Table \ref{tab_b5_b6}.  
Refer to Appendix, A3: Table \ref{tab_b7_b8} for the variations of $b_{7}$ and $b_{8}$.
\vspace{20pt}
\begin{center} {\Large \textbf {V. Conclusion}}\end{center}
{\large \textbf {a. AAB composition:}}

\indent Composite possible mathematical dependencies can finally be written [Equation (\ref{eq_rrm_aab}) and Equation (\ref{eq_critcomp_aab})] of IARRM (say, $\mu$) and $\Delta T$ for AAB configuration as:
\begin{eqnarray}
\label{eq_rrm_aab}
\left |\dfrac{M_{max,int}}{M_{sat}}\right|_{AAB}=\mu_{AAB}\left(\dfrac{J_{AA}}{J_{BB}},\dfrac{J_{AB}}{J_{BB}}\right)= a_{1}e^{-a_{2}|J_{AB}/J_{BB}|} \left[ a_{3}-a_{4}\left(\dfrac{J_{AA}}{J_{BB}}\right)^2 \right]\\
\label{eq_critcomp_aab}
\left[T_{crit}-T_{comp}\right]_{AAB}=\Delta T_{AAB}\left(\dfrac{J_{AA}}{J_{BB}},\dfrac{J_{AB}}{J_{BB}}\right)=\left[-a_{5}\left|\dfrac{J_{AB}}{J_{BB}}\right|+a_{6}\right]\left[-a_{7}\left(\dfrac{J_{AA}}{J_{BB}}\right)+a_{8}\right]
\end{eqnarray}
The coefficients $a_{1},a_{2},a_{3},a_{4},a_{5},a_{6},a_{7} \text{ and } a_{8}$ all are functions of $\dfrac{J_{AA}}{J_{BB}} \text{ and }\dfrac{J_{AB}}{J_{BB}}$.\\

\begin{figure}[!htb]
	\begin{center}
		\begin{tabular}{c}
			(a)
			\resizebox{9.25cm}{!}{\includegraphics[angle=0]{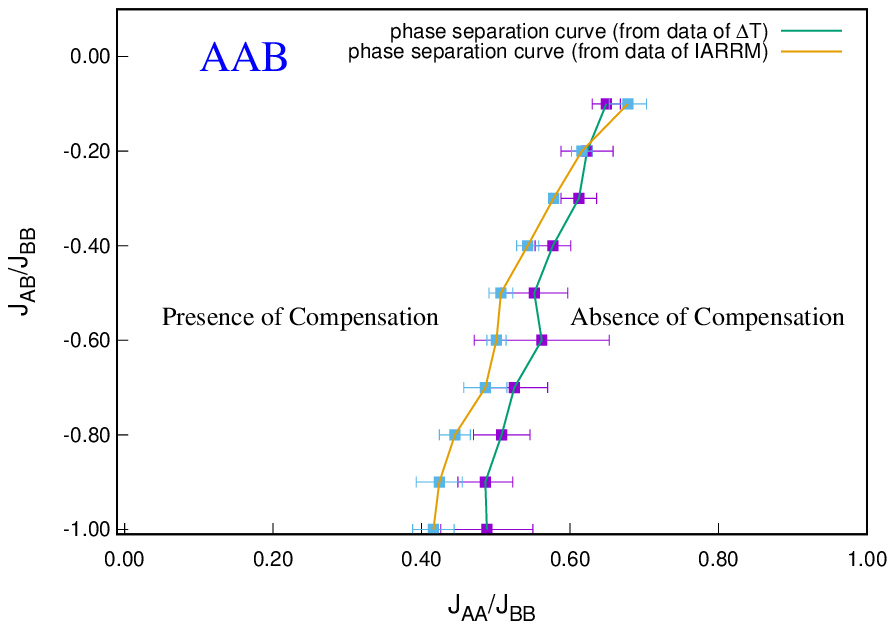}}
			(b)
			\resizebox{9.25cm}{!}{\includegraphics[angle=0]{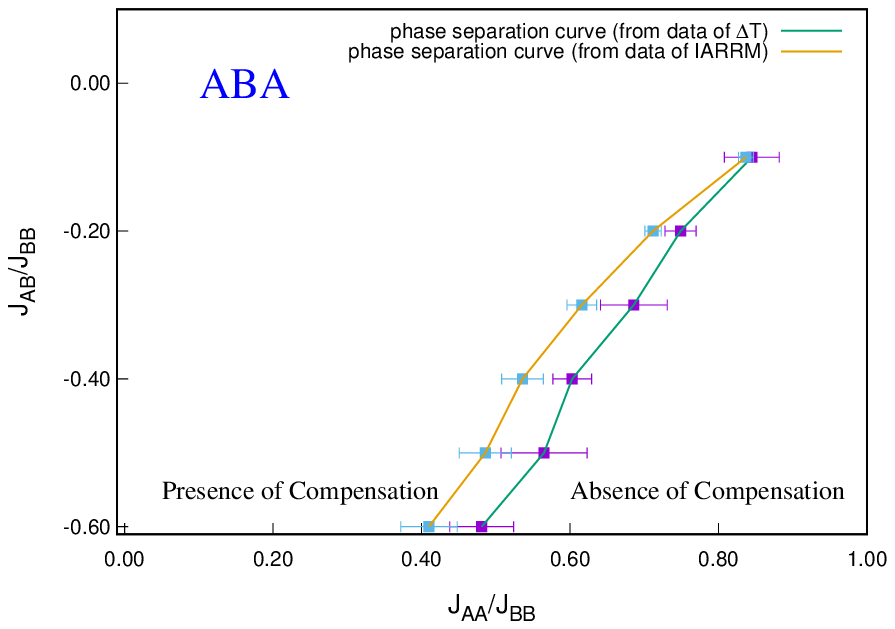}}
		\end{tabular}
		\caption{Phase diagrams of (a) AAB and (b) ABA configurations from the fitted formulae; in both of the figures: rightside of the curves denote ferrimagnetic phase without compensation and left of the curves denote ferrimagnetic phase with compensation.}		
		\label{fig_all_phasecurve}
	\end{center}
\end{figure}

\indent The zeroes of the Equation [\ref{eq_rrm_aab}] lead us to the specific combinations of the coupling ratios for AAB composition [Ref. Appendix, A2: Table \ref{tab_aab_rrm_phase2}] where the compensation ceases to exist within the ranges of our observation. Similarly zeroes of the Equation [\ref{eq_critcomp_aab}] were used to determine another set of specific combinations of the coupling ratios [Ref. Appendix, A2: Table \ref{tab_aab_critcomp_phase2}] for AAB configuration where the compensation just ceases to exist, for the data of temperature difference of $T_{crit}$ and $T_{comp}$. These combinations of coupling ratios in Tables \ref{tab_aab_rrm_phase2} \& \ref{tab_aab_critcomp_phase2} determine the phase curve of Figure \ref{fig_all_phasecurve}(a).\\
 
\vspace{10pt} 
\noindent{\large\textbf{b. ABA composition:}}
\vspace{5pt}
\\\indent Finally, possible composite mathematical dependencies are proposed [Equation (\ref{eq_rrm_aba}) and Equation (\ref{eq_critcomp_aba})] for IARRM (say, $\mu$) and $\Delta T$ for ABA configuration as:
\begin{eqnarray}
\label{eq_rrm_aba}
\left |\dfrac{M_{max,int}}{M_{sat}}\right|_{ABA}=\mu_{ABA}\left(\dfrac{J_{AA}}{J_{BB}},\dfrac{J_{AB}}{J_{BB}}\right)= b_{1}e^{-b_{2}(J_{AB}/J_{BB})^{2}} \left[ b_{3}-b_{4}\left(\dfrac{J_{AA}}{J_{BB}}\right)^2 \right]\\
\label{eq_critcomp_aba}
\left[T_{crit}-T_{comp}\right]_{ABA}=\Delta T_{ABA}\left(\dfrac{J_{AA}}{J_{BB}},\dfrac{J_{AB}}{J_{BB}}\right)=\left[-b_{5}\left|\dfrac{J_{AB}}{J_{BB}}\right|+b_{6}\right]\left[-b_{7}\left(\dfrac{J_{AA}}{J_{BB}}\right)+b_{8}\right]
\end{eqnarray}
The coefficients $b_{1},b_{2},b_{3},b_{4},b_{5},b_{6},b_{7} \text{ and } b_{8}$ all are functions of $\dfrac{J_{AA}}{J_{BB}} \text{ and }\dfrac{J_{AB}}{J_{BB}}$.\\
\\\indent As in the previous Section V(a), the zeroes of the Equation [\ref{eq_rrm_aba}] lead us to the specific combinations of the coupling ratios for IARRM data for ABA composition where the compensation just ceases to exist [Ref. Appendix, A4: Table \ref{tab_aba_rrm_phase}]. Similarly zeroes of Equation [\ref{eq_critcomp_aba}] lead us to the combinations of the coupling ratios for the $\Delta T$ data for ABA composition where the compensation just ceases to exist [Ref. Appendix, A4: Table \ref{tab_aba_critcomp_phase}]. These combinations of coupling ratios in Tables \ref{tab_aba_rrm_phase} \& \ref{tab_aba_critcomp_phase} determine the phase curve of Figure \ref{fig_all_phasecurve}(b). 
\\\indent Figure \ref{fig_all_phasecurve} shows the phase seperation curves obtained for two types of configurations under our ranges of observations which agree fairly well with \cite{Diaz4} where the curves are obtained via MFA, EFA and direct MC simulation data. If there are mathematical formulae, for determination of the presence and location of the compensation points, then this \textit{alternative description of the simulated systems helps technologists design materials according to their purpose without performing the entire range MC simulations at precise parameter values}. The layered magnetic materials, apt for compensation phenomenon, are economically cheap compared to rare-earth elements. Thus they are strong candidates for the MCE. 
\vspace{20pt}
\begin{center} {\Large \textbf {VI. Summary}}\end{center}
Summarising, magnetic and thermodynamic properties of a spin-1/2 Ising trilayered ferrimagnetic system on a square lattice are simulated. For an odd number of layers in layered ferrimagnets, neither of site dilution and mixed-spin cases is necessary for observing compensation on square trilayers. These types of systems are among the simplest of layered ferrimagnetic systems for compensation effect. Every layer of the bulk is composed of only one type of atoms out of two, i.e. A or B. The interactions between similar atoms (A-A or B-B bonds) are ferromagnetic and between dissimilar atoms (A-B) are antiferromagnetic, resulting in two distinct configurations: AAB and ABA. The Hamiltonian, used in this article, is defined with the help of the Ising interactions. Both the stackings are paramagnetic at high temperatures. The systems were initiated by an equal number of up $\left( S=+1 \right)$ and down $\left( S=-1 \right)$ spins, distributed randomly, mimicking a  paramagnetic phase with no magnetic order. Then the temperature was decreased in steps of $0.1$ till the final lowest temperature. Our objective is to find probable mathematical dependences of Inverse Absolute of Reduced Residual magnetisation and absolute temperature differences between $T_{crit}$ and $T_{comp}$, on relative interaction strengths in the Hamiltonian by analysing the system over a wide temperature range by MC single spin-flip algorithm and hence obtain the phase diagram along with the lattice morphology. MC method is simpler to understand, takes fluctuations into account contrary to MFA and yet produces accurate results. The findings are presented and analysed with the help of linear interpolation, and linear, quadratic and exponential curve fitting techniques.\\
\indent Morphological studies are not common in similar works thus initially, lattice morphologies were studied at $T_{crit}$ and in the neighbourhood of $T_{comp}$. The different spin density plots, of both the configurations, in those temperature regions [Fig.s \ref{fig_aab_morpho1}-\ref{fig_aab_morpho3} and Fig.s \ref{fig_aba_morpho1}-\ref{fig_aba_morpho3}] reveal that $T_{crit}$ and $T_{comp}$
are fundamentally different. The formation of spin clusters, of different sizes, at $T_{comp}$ is responsible for compensation phenomenon. The values of sublattice magnetizations are also provided which show them vanishing at $T_{crit}$ but at $T_{comp}$, non-zero values show the existence of magnetic ordering. Such studies help us understand the development of magnetic ordering in atomic or molecular level, while simulating through the entire range of operational temperaure.\\
\indent Next, for both AAB and ABA configurations, variations of Inverse Absolute of Reduced Residual Magnetisation (IARRM) and temperature gap between critical and compensation temperatures are presented. There are two parameters in the Hamiltonian, namely, $J_{AA}/J_{BB}$ and $J_{AB}/J_{BB}$ in the absence of an external magnetic field. $J_{AA}/J_{BB}$ are varied, from $0.1$ to $1.0$ and for each of the values $J_{AB}/J_{BB}$ ranges from $-0.1$ to $-1.0$. For each combination of them, the values of Inverse of Reduced Residual magnetisation and $|T_{crit}-T_{comp}|$ are obtained from the plots like Figure \ref{fig_aab_mag_fluc} and the 3D plots in Figures \ref{fig_aab_rrm_3d},\ref{fig_aab_critcomp_3d},\ref{fig_aba_rrm_3d},\ref{fig_aba_critcomp_3d}. Next, formulae for those mentioned quantities have been proposed. The systematics governed by these formulae lead to the phase diagrams (in the $J_{AB}/J_{BB}$-$J_{AA}/J_{BB}$ plane) for both AAB [Fig. \ref{fig_all_phasecurve}(a)] and ABA [Fig. \ref{fig_all_phasecurve}(b)] type systems, which divides the controlling parameter space into two regions: one with compensation and the other without compensation. In the phase diagrams where there is compensation, it can be seen that the range of $J_{AA}/J_{BB}$ increases as $|J_{AB}/J_{BB}|$ gets smaller and the compensation happens only when $J_{AA}<J_{BB}$ which is in fair agreement with \cite{Diaz4} and supports the findings in \cite{Diaz3, Balcerzak2, Szalowski} for similar systems made up of antiferromagnetic and ferromagnetic bonds. In the ABA configuration, the phase boundary is more inclined to the $J_{AB}/J_{BB}$ axis than the AAB configuration. In AAB configuration, the ratio of ferromagnetic to antiferromagnetic bonds per site is 5:1 for the A-layer in the middle and 4:1 for the bottom B-layer. In ABA configuration, compared to AAB, the ratio is 2:1 for the B-layer in the middle and 4:1 in the bottom A-layer. While in AAB, top A-layer has no antiferromagnetic bond, per site against a bond-ratio of 4:1 per site, in the top A-layer of ABA system. Greater number of antiferromagnetic bonds is responsible for the proneness to change of phase boundary with the change in the values of $J_{AB}/J_{BB}$ in ABA configuration.\\
\indent The technological realisation of a trilayered ferrimagnetic system with similar characteristics of the model in this article would be a great addition to the literature. 
The closed mathematical relations [Equation \ref{eq_rrm_aab}, \ref{eq_critcomp_aab}, \ref{eq_rrm_aba}, \ref{eq_critcomp_aba}] eliminates temeprature as an \textit{explicit} variable. \textit{For the systems presented here, being pure ones, the average magnetization at the lowest temperature ($\sim$ground state) for the systems are uniquely determined}. After determining the coupling constants \cite{Wolf} experimentally, and the extremum of average magnetisation (which happens at a finite, non-zero temperature, quite away from $0$K), one can determine the IARRM value for a specific combination of relative coupling strengths. Then with the help of the formulae for IARRM, one can obtain the range of values of either parameter (the relative coupling strengths: $J_{AA}/J_{BB}$ and $J_{AB}/J_{BB}$), where compensation is present (or absent) while keeping the other variable fixed. The similar argument extends for \textit{the temperature gap between compensation and critical temperatures}. Because of non-zero coupling between the mid and surface layers, the critical temperature for the entire system is the same as that of the mid-layer but here to detect the compensation temperature experimentally one has to go a bit lower in the temperature range than detecting the extremum of average magnetisation in the case of IARRM. Thus the extensive MC simulations or the study of entire temperature dependence of average magnetisation is not needed to be performed from the viewpoint of experimentalists, when following the presented formalism.
A recent study by Britton J.W. et al. \cite{Britton} shows experimental realisation of a triangular 2D Ising lattice with Beryllium ions stored in a Penning trap. Such engineered structures may lead to the creation of the studied systems on square lattices in future. In \cite{Chandra}, by extensive MC simulations, the comprehensive phase diagrams were obtained for triangular trilayered ferrimagnets of both: AAB and ABA configrations. In the context of magnetocaloric effects, investigation of entropy landscapes \cite{Jafari} in the vicinity of compensation and critical temperatures, for the systems in this article is also an interesting subject. Studies on entropy landscapes and the behaviours of IARRM and the absolute difference in the critical and compensation temperatures in the case of triangular trilayered ferrimagnet \cite{Chandra} and mixed spin \cite{Alzate1, Alzate2} systems, by MC simulational approach will be very interesting. These are planned for future. 
\vspace{20pt}
\begin{center} {\Large \textbf {Acknowledgements}}\end{center}
The author appreciates the financial assistance provided by University Grants Commission, India and gratefully acknowledges Prof. Muktish Acharyya for his comments and critical suggestions. The author also extends his thanks to his seniors Tamaghna Maitra and Sangita Bera for technical assistance. Several insightful comments and suggestions made by the anonymous referee are also acknowledged.
\vspace{50pt}
\begin{center} {\Large \textbf {References}} \end{center}

\newpage
\begin{center} {\huge \textbf {Appendix}}\end{center}
\vspace{20pt}
\begin{center} {\large \textbf {A1: Tables for different coefficients of AAB configuration}}\end{center}
\begin{table}[!htb]
	\centering
	\begin{tabular}{ c|c c c c c }
		\hline 
		\hline 
		$J_{AA}/J_{BB}$ & 0.1 & 0.2 & 0.3 & 0.4 & 0.5 \\
		\hline 
		$a_{1}$ & 0.966$\pm$0.007 & 0.944$\pm$0.016 & 0.940$\pm$0.016 & 0.855$\pm$0.027 & 0.818$\pm$0.049 \\
		\hline 
		$a_{2}$ & 0.890$\pm$0.014 & 1.166$\pm$0.039 & 1.713$\pm$0.045 & 2.662$\pm$0.109 & 5.336$\pm$0.344 \\
		\hline
		\hline  
	\end{tabular} 
	\caption{\label{tab_a1_a2}Variation of $a_{1}$ and $a_{2}$ with $J_{AA}/J_{BB}$, for AAB trilayer; the errors are standard asymptotic error for least square fitting}
\end{table} 
\begin{table}[!htb]
	\centering
	\begin{tabular}{ c|c c c c c c c c c c }
		\hline 
		\hline 
		$J_{AB}/J_{BB}$ & -0.1 & -0.2 & -0.3 & -0.4 & -0.5 & -0.6 & -0.7 & -0.8 & -0.9 & -1.0 \\
		\hline 
		$a_{3}$ & 0.945 & 0.852 & 0.761 & 0.674 & 0.634 & 0.562 & 0.497 & 0.487 & 0.443 & 0.401\\
		& $\pm$0.027 & $\pm$0.017 & $\pm$0.003 & $\pm$0.016 & $\pm$0.020 & $\pm$0.014 & $\pm$0.016 & $\pm$0.031 & $\pm$0.030 & $\pm$0.026 \\
		\hline 
		$a_{4}$ & 2.055 & 2.241 & 2.281 & 2.289 & 2.470 & 2.241 & 2.104 & 2.461 & 2.469 & 2.314 \\
		& $\pm$0.138 & $\pm$0.088 & $\pm$0.022 & $\pm$0.118 & $\pm$0.140 & $\pm$0.101 & $\pm$0.241 & $\pm$0.168 & $\pm$0.320 & $\pm$0.279 \\
		\hline
		\hline  
	\end{tabular} 
	\caption{\label{tab_a3_a4}Variation of $a_{3}$ and $a_{4}$ with $J_{AB}/J_{BB}$, for AAB trilayer; the errors are standard asymptotic errors for least square fitting}
\end{table} 
\begin{table}[!htb]
	\centering\offinterlineskip
	\begin{tabular}{ c|c c c c c }
		\hline 
		\hline 
		$J_{AA}/J_{BB}$ & 0.1 & 0.2 & 0.3 & 0.4 & 0.5 \\
		\hline 
		$a_{5}$ & 0.476$\pm$0.026 & 0.648$\pm$0.036 & 0.587$\pm$0.039 & 0.508$\pm$0.072 & 0.407$\pm$0.113 \\
		\hline 
		$a_{6}$ & 1.876$\pm$0.016 & 1.563$\pm$0.022 & 1.164$\pm$0.024 & 0.793$\pm$0.044 & 0.497$\pm$0.051 \\
		\hline
		\hline  
	\end{tabular} 
	\caption{\label{tab_a5_a6}Variation of $a_{5}$ and $a_{6}$ with FM ratio, for AAB trilayer; the errors are standard asymptotic error for least square fitting}%
\end{table} 
\begin{table}[!htb]
	\centering\offinterlineskip 
	\begin{tabular}{ c|c c c c c c c c c c }
		\hline 
		\hline 
		$J_{AB}/J_{BB}$ & -0.1 & -0.2 & -0.3 & -0.4 & -0.5 & -0.6 & -0.7 & -0.8 & -0.9 & -1.0 \\
		\hline 
		$a_{7}$ & 3.308 & 3.346 & 3.240 & 
		3.540 & 3.560 & 3.280 & 3.490 & 
		3.550 & 3.620 & 3.420\\
		& $\pm$0.082 & $\pm$0.160 & $\pm$0.111 & $\pm$0.128 & $\pm$0.250 & $\pm$0.457 & $\pm$0.251 & $\pm$0.231 & $\pm$0.242 & $\pm$0.380 \\
		\hline 
		$a_{8}$ & 2.148 & 2.086 & 1.984 &
		2.044 & 1.964 & 1.842 & 1.831 & 
		1.805 & 1.760 & 1.670 \\
		& $\pm$0.032 & $\pm$0.062 & $\pm$0.037 & $\pm$0.042 & $\pm$0.083 & $\pm$0.152 & $\pm$0.083 & $\pm$0.063 & $\pm$0.066 & $\pm$0.104 \\
		\hline
		\hline  
	\end{tabular} 
	\caption{\label{tab_a7_a8}Variation of $a_{7}$ and $a_{8}$ with AFM ratio, for AAB trilayer; the errors are standard asymptotic errors for least square fitting}
\end{table}
\begin{center} {\large \textbf {A2: Tables for Phase diagram of AAB configuration}}\end{center}
\begin{table}[!htb]
	\centering\offinterlineskip 
	\begin{tabular}{ c|c c c c c c c c c c}
		\hline 
		\hline 
		$J_{AB}/J_{BB}$ & -0.1 & -0.2 & -0.3 & -0.4 & -0.5 & -0.6 & -0.7 & -0.8 & -0.9 & -1.0 \\
		\hline 
		$J_{AA}/J_{BB}$ & 0.678 & 0.616 & 0.578 & 0.543 & 0.507 & 0.501 & 0.486 & 0.445 & 0.424 & 0.416 \\
		& $\pm$0.008 & $\pm$0.014 & $\pm$0.017 & $\pm$0.021 & $\pm$0.024 & $\pm$0.024 & $\pm$0.038 & $\pm$0.025 & $\pm$0.035 & $\pm$0.034 \\
		\hline
		\hline  
	\end{tabular} 
	\caption{\label{tab_aab_rrm_phase2}Table for maximum values of $J_{AA}/J_{BB}$ for a fixed $J_{AB}/J_{BB}$ for which the compensation effect just ceases, from the fitted formula of variation of IARRM of AAB configuration.}
\end{table}

\begin{table}[!htb]
	\centering\offinterlineskip 
	\begin{tabular}{ c|c c c c c c c c c c}
		\hline 
		\hline 
		$J_{AB}/J_{BB}$ & -0.1 & -0.2 & -0.3 & -0.4 & -0.5 & -0.6 & -0.7 & -0.8 & -0.9 & -1.0 \\
		\hline 
		$J_{AA}/J_{BB}$ & 0.649 & 0.623 & 0.612 & 0.577 & 0.552 & 0.562 & 0.525 & 0.508 & 0.486 & 0.488 \\
		& $\pm$0.019 & $\pm$0.035 & $\pm$0.024 & $\pm$0.024 & $\pm$0.045 & $\pm$0.091 & $\pm$0.045 & $\pm$0.038 & $\pm$0.037 & $\pm$0.062 \\
		\hline
		\hline  
	\end{tabular} 
	\caption{\label{tab_aab_critcomp_phase2}Maximum values of $J_{AA}/J_{BB}$ for fixed $J_{AB}/J_{BB}$ for which the compensation effect just ceases, from the fitted formula of $\Delta T$ of AAB configuration.}
\end{table}
\newpage 
\begin{center} {\large \textbf {A3: Tables for different coefficients of ABA configuration}}\end{center}
\begin{table}[!htb]
	\centering
	\begin{tabular}{ c|c c c c c }
		\hline 
		\hline 
		$J_{AA}/J_{BB}$ & 0.1 & 0.2 & 0.3 & 0.4 & 0.5 \\
		\hline 
		$b_{1}$ & 0.791$\pm$0.020 & 0.770$\pm$0.020 & 0.733$\pm$0.026 & 0.699$\pm$0.024 & 0.635$\pm$0.026 \\
		\hline 
		$b_{2}$ & 2.804$\pm$0.155 & 4.048$\pm$0.229 & 5.722$\pm$0.430 & 8.972$\pm$0.611 & 14.554$\pm$1.14 \\
		\hline
		\hline  
	\end{tabular} 
	\caption{\label{tab_b1_b2}Variation of $b_{1}$ and $b_{2}$ with $J_{AA}/J_{BB}$, for ABA trilayer; the errors are standard asymptotic error for least square fitting}
\end{table} 
\begin{table}[!htb]
	\centering
	\begin{tabular}{ c|c c c c c c }
		\hline 
		\hline 
		$J_{AB}/J_{BB}$ & -0.1 & -0.2 & -0.3 & -0.4 & -0.5 & -0.6 \\
		\hline 
		$b_{3}$ & 0.841$\pm$0.009 & 0.682$\pm$0.009 & 0.562$\pm$0.017 & 0.464$\pm$0.021 & 0.363$\pm$0.026 & 0.290$\pm$0.026 \\
		\hline 
		$b_{4}$ & 1.201$\pm$0.027 & 1.345$\pm$0.036 & 1.482$\pm$0.086 & 1.615$\pm$0.152 & 1.540$\pm$0.189 & 1.729$\pm$0.282 \\
		\hline
		\hline  
	\end{tabular} 
	\caption{\label{tab_b3_b4}Variation of $b_{3}$ and $b_{4}$ with $J_{AB}/J_{BB}$, for ABA trilayer; the errors are standard asymptotic error for least square fitting}
\end{table} 
\begin{table}[!htb]
	\centering\offinterlineskip
	\begin{tabular}{ c|c c c c c }
		\hline 
		\hline 
		$J_{AA}/J_{BB}$ & 0.1 & 0.2 & 0.3 & 0.4 & 0.5 \\
		\hline 
		$b_{5}$ & 0.476$\pm$0.026 & 0.648$\pm$0.036 & 0.587$\pm$0.039 & 0.508$\pm$0.072 & 0.407$\pm$0.113 \\
		\hline 
		$b_{6}$ & 1.876$\pm$0.016 & 1.563$\pm$0.022 & 1.164$\pm$0.024 & 0.793$\pm$0.045 & 0.497$\pm$0.051 \\
		\hline
		\hline  
	\end{tabular} 
	\caption{\label{tab_b5_b6}Variation of $b_{5}$ and $b_{6}$ with $J_{AA}/J_{BB}$, for ABA trilayer; the errors are standard asymptotic error for least square fitting}%
\end{table} 
\begin{table}[!htb]
	\centering\offinterlineskip 
	\begin{tabular}{ c|c c c c c c }
		\hline 
		\hline 
		$J_{AB}/J_{BB}$ & -0.1 & -0.2 & -0.3 & -0.4 & -0.5 & -0.6 \\
		\hline 
		$b_{7}$ & 2.433$\pm$0.091 & 2.464$\pm$0.059 & 2.506$\pm$0.142 & 2.470$\pm$0.093 & 2.370$\pm$0.208 & 2.430$\pm$0.190\\
		\hline 
		$b_{8}$ & 2.056$\pm$0.046 & 1.846$\pm$0.026 & 1.719$\pm$0.055 & 1.489$\pm$0.031 & 1.339$\pm$0.069 & 1.170$\pm$0.052\\
		\hline
		\hline  
	\end{tabular} 
	\caption{\label{tab_b7_b8}Variation of $b_{7}$ and $b_{8}$ with $J_{AB}/J_{BB}$, for ABA trilayer; the errors are standard asymptotic errors for least square fitting}
\end{table}
\vspace{20pt}
\begin{center} {\large \textbf {A4: Tables for Phase diagram of ABA configuration}}\end{center}
\begin{table}[!htb]
	\centering\offinterlineskip 
	\begin{tabular}{ c|c c c c c c }
		\hline 
		\hline 
		$J_{AB}/J_{BB}$ & -0.1 & -0.2 & -0.3 & -0.4 & -0.5 & -0.6 \\
		\hline 
		$J_{AA}/J_{BB}$ & 0.837$\pm$0.010 & 0.712$\pm$0.011 & 0.616$\pm$0.020 & 0.536$\pm$0.028 & 0.486$\pm$0.035 & 0.410$\pm$0.038 \\
		\hline
		\hline  
	\end{tabular} 
	\caption{\label{tab_aba_rrm_phase}Table for maximum values of $J_{AA}/J_{BB}$ for a fixed $J_{AB}/J_{BB}$ for which the compensation effect just ceases, from the fitted formula of variation of IARRM of ABA configuration.}
\end{table}
\begin{table}[!htb]
	\centering\offinterlineskip 
	\begin{tabular}{ c|c c c c c c }
		\hline 
		\hline 
		$J_{AB}/J_{BB}$ & -0.1 & -0.2 & -0.3 & -0.4 & -0.5 & -0.6 \\
		\hline 
		$J_{AA}/J_{BB}$ & 0.845$\pm$0.037 & 0.749$\pm$0.021 & 0.686$\pm$0.045 & 0.603$\pm$0.026 & 0.565$\pm$0.058 & 0.481$\pm$0.043 \\ 
		\hline
		\hline  
	\end{tabular} 
	\caption{\label{tab_aba_critcomp_phase}Maximum values of $J_{AA}/J_{BB}$ for a fixed $J_{AB}/J_{BB}$ for which the compensation effect just ceases, from the fitted formula of variation of $\Delta T$ of ABA configration.}
\end{table}

\end{document}